\documentclass[letterpaper,twocolumn,10pt]{article}
\usepackage{usenix-2020-09}
\usepackage{xspace}

\newcommand{\AUTHORS}{Authors}

\newcommand{\NAME}{MoEvement\xspace}

\newcommand{\NAMESC}{\textsc{MoEvement}\xspace}

\newcommand{\TITLE}{Sparse Checkpointing for Fast and Reliable MoE Training}

\newcommand{\KEYWORDS}{}
\newcommand{\CONFERENCE}{}
\newcommand{\PAGENUMBERS}{yes}  
\newcommand{\COMMENTS}{no}      

\newcommand{\ie}{{i.e.,}~}
\newcommand{\eg}{{e.g.,}~}

\newcommand{\paraf}[1]{\noindent\textbf{#1}}

\usepackage{balance, ifthen, multirow, times}

\usepackage{inconsolata}

\usepackage{titling}            
\usepackage[small,compact]{titlesec}         

\usepackage[font=small, labelfont=bf]{caption}

\usepackage{subcaption}

\usepackage{fancyhdr}

\ifthenelse{\equal{\PAGENUMBERS}{yes}}{%
  \pagestyle{plain}
}{%
  \pagestyle{empty}
}

\usepackage[numbers,sort&compress]{natbib}

\usepackage{enumitem}
\setlist{itemsep=0pt,parsep=0pt,topsep=0pt} 

\usepackage{booktabs}
\usepackage{colortbl}
\usepackage{float}                           

\newcommand*{\thinline}{\specialrule{.01em}{0.05em}{0.05em}}

\usepackage{pgfplots}
\usepackage{array}
\usepackage{graphicx}
\pgfplotsset{compat=1.18}

\newcommand*{\MinNumber}{0.1}%
\newcommand*{\MidNumber}{0.85} %
\newcommand*{\MaxNumber}{1.0}%

\newcommand{\ApplyGradient}[1]{%
        \ifdim #1 pt > \MidNumber pt
            \pgfmathsetmacro{\PercentColor}{max(min(100.0*(#1 - \MidNumber)/(\MaxNumber-\MidNumber),100.0),0.00)} %
            \hspace{-0.33em}\colorbox{green!\PercentColor!yellow}{#1}
        \else
            \pgfmathsetmacro{\PercentColor}{max(min(100.0*(\MidNumber - #1)/(\MidNumber-\MinNumber),100.0),0.00)} %
            \ifdim #1 pt < 0.65 pt 
                \hspace{-0.33em}\colorbox{red!\PercentColor!yellow}{\color{white}{#1}}
            \else
                \hspace{-0.33em}\colorbox{red!\PercentColor!yellow}{#1}
            \fi
        \fi
}

\newcommand{\Best}[1]{%
  \hspace{-0.33em}%
  \colorbox{green!75!yellow}{\makebox[1.65em][c]{\textbf{#1}}}%
}
\newcommand{\Worst}[1]{%
  \hspace{-0.33em}%
  \colorbox{red!100!yellow}{\color{white}{\makebox[1.65em][c]{\textbf{#1}}}}%
}

\usepackage{makecell}
\usepackage[dvipsnames]{xcolor}
\usepackage{pifont}
\newcommand{\greencheck}{\color{ForestGreen}{\pmb{\ding{51}}}}
\newcommand{\redcross}{\color{red}{\pmb{\ding{55}}}}

\usepackage{mathtools}
\usepackage{amssymb}
\usepackage{pifont}

\usepackage{url}
\hypersetup{%
    pdfauthor = {\AUTHORS},
    pdftitle = {\TITLE},
    pdfsubject = {\CONFERENCE},
    pdfkeywords = {\KEYWORDS},
    bookmarksopen = {true}
}
\usepackage{cleveref}               
\crefname{section}{\S}{\SS}
\crefformat{section}{\S#2#1#3}      
\crefformat{subsection}{\S#2#1#3}
\crefformat{subsubsection}{\S#2#1#3}

\usepackage{listings}
\newcommand\code[1]{\lstinline$#1$}

\DeclareRobustCommand{\bigO}{%
  \text{\usefont{OMS}{cmsy}{m}{n}O}%
}

\usepackage{algorithm}
\usepackage{algpseudocode}
\usepackage{multirow}
\usepackage{pifont}
\usepackage[normalem]{ulem}

\definecolor{string-color}{rgb}{0.3333, 0.5254, 0.345}
\definecolor{commentcolor}{HTML}{5d6063}
\lstdefinestyle{pseudocode}{
    language=Python,
    mathescape=true,
    keywords={FindWindowSize, GenerateSchedule, OrderOperators, SparseCheckpointSchedule, profiler},
    keywordstyle=\color{blue}\bfseries,
    morekeywords={[2]},
    keywordstyle={[2]\bfseries},
    morekeywords={[3]if,def,return,else,while,for,in,append,ceil,break},
    keywordstyle={[3]\bfseries},
    basicstyle=\small\ttfamily,
    identifierstyle=\color{black},
    sensitive=false,
    commentstyle=\color{commentcolor},
    columns=fullflexible,
    keepspaces=false,
    breaklines=true,
    showstringspaces=false,
    numbers=left,
    numberstyle=\tiny,
    numbersep=4pt,
    xleftmargin=5pt,
    lineskip={-1pt},
    postbreak=\mbox{\textcolor{black}{$\hookrightarrow$}\space},
    stringstyle=\color{string-color},
}

\makeatletter
\newenvironment{codealgorithm}[1][htb]{%
  \renewcommand{\ALG@name}{Algorithm}%
  \begin{algorithm}[#1]%
  }{\end{algorithm}}
\makeatother

\algnewcommand{\LineComment}[1]{\State \(\Comment\) #1}

\usepackage{tikz}
\DeclareRobustCommand\numcircledtikz[1]{\tikz[baseline=(char.base)]{
    \node[shape=circle,draw,fill,inner sep=1pt] (char)
    {\textcolor{white}{#1}};}}

\DeclareRobustCommand\greycircledtikz[1]{%
  \tikz[baseline=(char.base)]{
    \node[shape=circle,draw=gray!75,fill=gray!90,inner sep=0.75pt] (char)
    {\textcolor{white}{#1}};}}

\usepackage{comment} 
\usepackage{soul}
\soulregister\cite7
\soulregister\ref7
\soulregister\pageref7

\usepackage[all]{nowidow}

\ifthenelse{\equal{\COMMENTS}{yes}}{%
    \newcommand\swapnil[1]{\textcolor{blue}{[Swapnil:] #1}}
    \newcommand\christos[1]{\textcolor{magenta}{[Christos:] #1}}
}{
    \newcommand\swapnil[1]{\unskip}
    \newcommand\christos[1]{\unskip}
}

\begin{document}
\date{}

\title{\bf \TITLE}

\author{
{\rm Swapnil Gandhi}\\
Stanford University\\
\and
{\rm Christos Kozyrakis}\\
Stanford University \& NVIDIA\\
}

\maketitle  

\ifthenelse{\equal{\PAGENUMBERS}{no}}{%
  \thispagestyle{empty}
}

\makeatletter
\def\blfootnote{\xdef\@thefnmark{}\@footnotetext}
\makeatother

\begin{abstract}
As large language models scale, training them requires thousands of GPUs over extended durations—making frequent failures an inevitable reality. While checkpointing remains the primary fault-tolerance mechanism, existing methods fall short when applied to Mixture-of-Experts (MoE) models. Due to their substantially larger training state, MoE models exacerbate checkpointing overheads, often causing costly stalls or prolonged recovery that severely degrade training efficiency.

We present MoEvement, a distributed, in-memory checkpointing system tailored for MoE models. MoEvement is built on three key ideas: (1) \emph{sparse checkpointing}, which incrementally snapshots subsets of experts across iterations to reduce overhead; (2) a \emph{sparse-to-dense checkpoint conversion} mechanism that incrementally reconstructs consistent dense checkpoints from sparse snapshots; and (3) \emph{upstream logging} of activations and gradients at pipeline-stage boundaries, enabling localized recovery without re-executing unaffected workers. Evaluations across diverse MoE models with up to 64 experts show that MoEvement reduces checkpointing overhead by up to \(4\times\) and recovery overhead by up to \(31\times\) compared to state-of-the-art approaches, sustaining ETTR $\ge 0.94$ even under frequent failures (MTBF as low as 10 minutes) and delivering up to $8\times$ overall training speedup, all without compromising synchronous training semantics. Overall, MoEvement offers a robust and scalable fault-tolerance solution for the next generation of sparsely activated models.
\end{abstract}

\section{Introduction}\label{sec:introduction}

The rise of large foundation models has driven orders-of-magnitude growth in distributed training infrastructure~\cite{megatron-sc, pathways, deepspeed, zero, zero-infinity, pytorch-distributed, pytorch-fsdp, deepseek-r1}. Training frontier models, such as DeepSeek-V3 with 671B parameters trained on 14.8 trillion tokens~\cite{deepseek-v3}, requires thousands of GPUs over extended periods~\cite{chinchilla, training-at-meta, megascale, bloom, gpt-4, nemotron-moe}. At these scales, system failures—triggered by hardware faults, network disruptions, or software bugs—shift from being rare anomalies to frequent events. Major organizations including Microsoft, ByteDance, Alibaba, and Google have reported failures as frequently as once every 45 minutes, with the mean time between failures (MTBF) decreasing as GPU count rises~\cite{failures-in-large-scale-systems, megascale, unicorn, tpu-resiliency, reliability-at-scale-meta}. Meta projects an MTBF as low as 14 minutes for jobs utilizing 131,072 GPUs~\cite{reliability-at-scale-meta}, underscoring an emerging reality: \textit{at scale, failure is not an anomaly—it is the norm}.

\begin{figure*}[t!]
    \centering 
    \begin{subfigure}[t]{0.49\linewidth}
        \centering
        \captionsetup{justification=centering}
        \includegraphics[width=\linewidth]{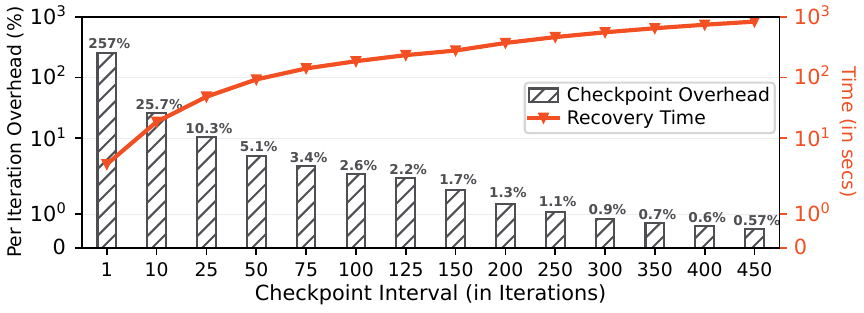}
        \caption{Checkpoint interval vs. per-iteration \% overhead (bar, log-scale, left Y-axis) and recovery time (line, log-scale, right Y-axis)}
        \label{fig:checkpoiniting-overheads}
    \end{subfigure} 
    \hfill
    \begin{subfigure}[t]{0.49\linewidth}
        \centering
        \captionsetup{justification=centering}
        \includegraphics[width=\linewidth]{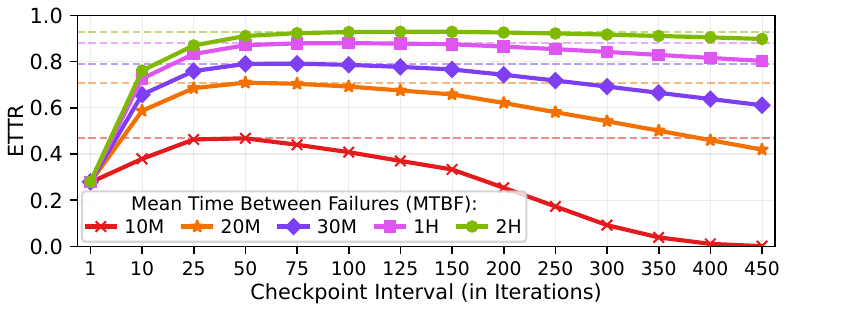}
        \caption{ETTR across checkpoint intervals for varying MTBFs. Dashed lines indicate the maximum ETTR achieved for each MTBF.}
        \label{fig:ettr-mtbf}
    \end{subfigure}
    \vspace{-0.075in} 
    \caption{Performance of Gemini~\cite{gemini} during training of DeepSeek-16.4B/64-Experts MoE model~\cite{deepseek-moe} using 96 A100 GPUs.} 
    \label{fig:dense-checkpointing-perf-comparision}
    \vspace{-0.1in} 
\end{figure*}

Checkpointing—periodically capturing model and optimizer states for durable persistence—is the standard fault-tolerance mechanism to limit computation lost to failures~\cite{failures-in-large-scale-systems, bloom-chronicles, megascale, unicorn, llama-4, gpt-4, deepseek-r1}. To reduce checkpoint overhead, recent techniques leverage overlapping snapshot operations with computation~\cite{checkfreq, bytecheckpoint, lazy-checkpoint}, high-performance interconnects~\cite{fast-persist, pccheck, gemini}, checkpoint compression~\cite{inshrinkerator, check-n-run, cpr}, and redundancy across data-parallel nodes~\cite{just-in-time}. However, these approaches primarily target dense models, where all parameters contribute uniformly to computation and convergence.

In contrast, Mixture-of-Experts (MoE) models follow a fundamentally different paradigm—replacing dense layers with tens to hundreds (occasionally millions~\cite{peer}) of sparsely activated expert operators~\cite{switch-transformer, glam, deepseek-moe, mixtral}. Each token activates only a few experts (commonly four or eight out of hundreds~\cite{moe-survey}), enabling an order of magnitude larger parameter counts without proportional increases in computation~\cite{scaling-laws-moe, gshard}. This sparse, dynamically varying operator activation pattern presents fundamental challenges for traditional checkpointing techniques. In particular, MoE models expose three key limitations in existing approaches:

{\bf Challenge \#1: Runtime–Recovery Tradeoff:}
State-of-the-art checkpointing techniques—such as CheckFreq~\cite{checkfreq}, which pipelines the \textit{snapshot} (copying state to local CPU memory) and \textit{persist} (flushing to durable storage) phases, and Gemini~\cite{gemini}, which uses high-bandwidth remote CPU memory to overlap snapshot I/O with computation—while effective for dense models, fall short for MoE models, which expand training state by an order of magnitude without increasing iteration times. This expansion makes short checkpoint intervals prohibitively expensive: checkpointing every iteration for DeepSeek-16.4B/64E slows training by $2.5\times$ under Gemini (Fig.~\ref{fig:checkpoiniting-overheads}, left Y-axis). Increasing the interval reduces per-iteration checkpoint overhead but lengthens recovery time (Fig.~\ref{fig:checkpoiniting-overheads}, right Y-axis), as the average recomputation after a failure grows with interval length~\cite{daly-2006}. When failures are infrequent, this tradeoff has limited impact, but as MTBF drops—such as in large-scale runs where failures occur frequently—recovery costs compound rapidly, driving down ETTR\footnote{Effective Training Time Ratio (ETTR) is the fraction of wall‑clock time spent on useful training, excluding checkpointing and recovery.~\cite{reliability-at-scale-meta}.}. As shown in Figure~\ref{fig:ettr-mtbf}, Gemini’s ETTR peaks at $0.93$ (2-hour MTBF) but is limited to at most $0.79$ at 30-minute MTBF, plunging further to $0.47$ at 10-minute MTBF. At million-dollar training budgets~\cite{rising-cost-2024}, such drops translate to hundreds of thousands of dollars in wasted computation, making these inefficiencies prohibitive at scale.

In this paper, we introduce \NAME\footnote{Pronounced ``movement,'' referring to its focus on MoE models and the movement of training state to ensure continued progress despite failures.}, which breaks the runtime–recovery tradeoff via \emph{sparse checkpointing} (\cref{sec:sparse-checkpointing}). Instead of checkpointing the full training state in a single iteration, \NAME incrementally snapshots subsets of operators across multiple iterations, evenly distributing and fully overlapping checkpointing I/O with computation. By prioritizing operators based on activation frequency (\cref{sec:checkpointing-policy}), \NAME enables low-overhead, high-frequency checkpoints. This removes the need to choose between frequent-but-expensive and infrequent-but-costly-to-recover checkpoints, allowing \NAME to sustain ETTR \(\ge0.94\), even at low MTBFs.

{\bf Challenge \#2: Correctness–Efficiency Tension:} 
Recent MoE-specific checkpointing approaches, like MoC-System~\cite{moc-system}, attempt to reduce checkpoint overhead by snapshotting only a subset of experts per iteration in a round-robin fashion. While this lowers the overhead of frequent checkpoints, it compromises correctness: during recovery, experts without recent checkpoints revert to stale parameters, causing token loss and violating synchronous training semantics~\cite{convergence-analysis}. 

\NAME overcomes the correctness–efficiency trade-off with a \emph{sparse-to-dense checkpoint conversion} mechanism (\cref{sec:sparse-to-dense-conversion}). Each sparse checkpoint stores full FP32 state for a subset of operators and FP16 state for the rest\footnote{Unless stated otherwise, and following standard practice~\cite{megatron-sc, gopher}, we assume mixed-precision training: FP32 (32-bit floating point) master weights and optimizer states ensure numerical stability, while FP16 weights are used for forward and backward computation. As we show in~\cref{sec:low-precision-exp}, our techniques are also applicable to low-precision training regimes, including FP8 formats.}. During recovery, operators are incrementally restored from FP16 to FP32, with iterations recomputed as needed until a consistent dense checkpoint is reconstructed. \NAME preserves synchronous semantics, maintains accuracy, and retains reproducibility, ensuring no loss of tokens or training progress.

{\bf Challenge \#3: Global Rollback Scope:}  
To ensure synchronous semantics, existing checkpointing techniques roll back all workers, faulty or not, to a common checkpoint~\cite{checkfreq, gemini, just-in-time, bytecheckpoint, pccheck}. This global rollback amplifies recomputation overhead and prolongs recovery, a cost that grows sharply at scale, where a single worker failure can force hundreds of otherwise healthy workers to revert training progress.

\NAME employs \emph{Upstream Logging} (\cref{sec:upstream-logging}), a targeted recovery mechanism that narrows rollback scope by logging intermediate activations and gradients at pipeline stage boundaries. On failure, only the affected data-parallel group rolls back to its most recent sparse checkpoint—typically just a few iterations—and completes \emph{sparse-to-dense conversion} directly from the stored logs. This localized recovery shortens recovery time and eliminates the global rollback overhead.

We implemented \NAME on top of DeepSpeed~\cite{deepspeed} and evaluated it across diverse MoE models spanning both vision and language domains, where it consistently sustains ETTR \(\ge 0.94\) even under frequent failures, delivering up to $8\times$ overall training speedup. These gains stem from reducing checkpointing overhead by up to $4\times$ compared to MoC-System~\cite{moc-system} and accelerating recovery by up to $31\times$ and $17\times$ relative to CheckFreq~\cite{checkfreq} and Gemini~\cite{gemini}, respectively. Crucially, these efficiency improvements come with no loss in model accuracy. By breaking the runtime–recovery tradeoff, resolving correctness–efficiency tensions in prior checkpointing methods, and eliminating unnecessary global rollbacks, \NAME provides a robust, scalable fault-tolerance solution purpose-built for large-scale MoE training.
\section{Background \& Related Work}\label{sec:background}

\subsection{Sparse Mixture-of-Experts Models}
Mixture-of-Experts (MoE) models have emerged as a scalable and compute-efficient architecture for training large-scale neural networks across various domains. MoEs extend standard transformer by replacing dense feed-forward layers with multiple parallel subnetworks, known as \textit{experts}~\cite{shazeer2017}, each activated selectively based on the input. A learned gating network routes each token to a small subset of these experts—typically one or two—and combines their outputs through learned weights. This selective activation enables MoE models to scale total parameter counts with only sub-linear growth in computational overhead, underpinning state-of-the-art foundation models such as DeepSeek-V3~\cite{deepseek-v3}, Gemini~\cite{ggemini}, Grok~\cite{grok}, gpt-oss~\cite{gpt-oss}, and Llama 4~\cite{llama-4}.

To mitigate expert imbalance, MoEs commonly use auxiliary load-balancing objectives~\cite{switch-transformer, demons-in-detail, st-moe} to encourage more uniform expert activation. In practice, however, activations remains naturally skewed due to input diversity, expert specialization, and training dynamics~\cite{smartmoe, fastermoe, fastmoe}. Attempts at strict balancing through strong regularization can disrupt expert specialization and degrade model performance~\cite{deepseek-v3, switch-transformer}.

\begin{figure}[t!]
    \centering
    \includegraphics[width=1.0\linewidth]{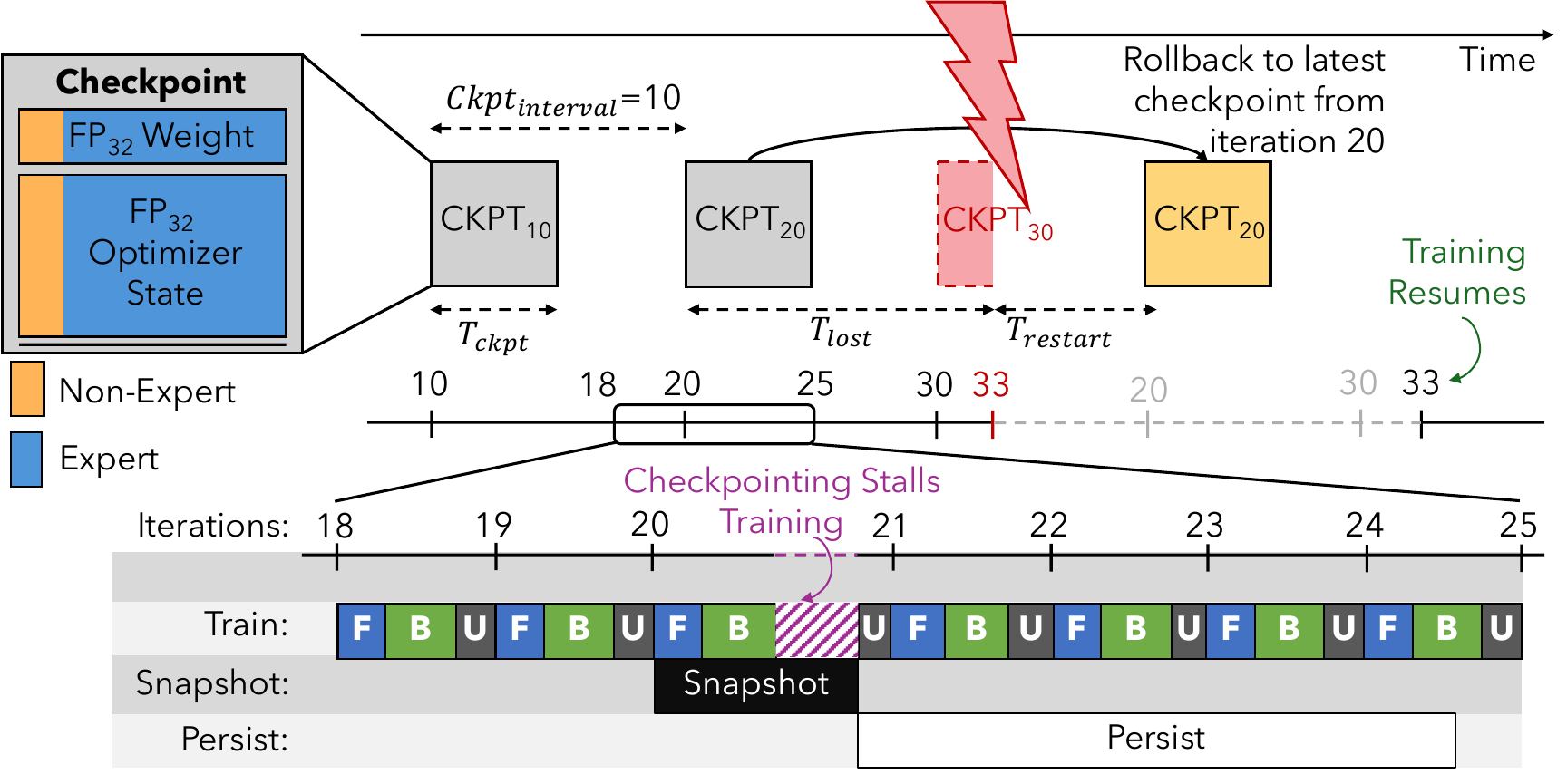}
    \vspace{-0.25in} 
    \caption{Checkpoint-based fault tolerance in distributed training. A checkpoint is taken every \( Ckpt_{\text{interval}} = 10 \) iterations. On failure, training rolls back to the most recent complete checkpoint (\( \text{CKPT}_{20} \)) and recomputes lost training progress, and then resumes.}
    \label{fig:checkpoint-flow}
\end{figure}

\subsection{Distributed Training of MoE Models}

State-of-the-art MoE models contain hundreds of billions of parameters trained on datasets with trillions of tokens~\cite{chinchilla, megatron-lm, switch-transformer, llama-3}. Training typically spans weeks and requires thousands of GPUs~\cite{llama-4, megascale, pathways}. For instance, DeepSeek-V3, a 671B-parameter model, trained on 14.8 trillion tokens, consumed 2.7 million Nvidia H800 GPU hours—equivalent to 11 days on a 10,000-GPU cluster~\cite{deepseek-v3}.

MoE training employs four primary parallelism strategies. \emph{Data parallelism} (DP) splits batches across GPUs, synchronizing parameters with all-reduce operations~\cite{horovod, imagenet-in-hours}. \emph{Tensor parallelism} (TP) partitions model layers across GPUs to accelerate large matrix operations (e.g., GEMMs)~\cite{megatron-lm}. \emph{Pipeline parallelism} (PP) divides the model sequentially, processing micro batches in a pipelined fashion~\cite{gpipe, pipedream, chimera, bpipe}. \emph{Expert parallelism} (EP), unique to MoEs, distributes experts across GPUs and routes tokens using all-to-all communication, enabling scalable conditional computation~\cite{gshard, fastmoe, smartmoe, tutel}.

\paragraph{Frequent Failures in Distributed Training.} Modern training clusters comprise thousands of GPUs interconnected by sophisticated networking, storage, and power systems~\cite{reliability-at-scale-meta}. 
At this scale, failures are inevitable, arising from hardware faults (e.g., overheating GPUs, SSD degradation), transient network disruptions, software crashes, or service-level issues (e.g., monitoring agents exceeding resource limits)~\cite{tpu-resiliency, megascale, failure-characterization-in-datacenter}. Alibaba Cloud reported abnormal termination rates of 44\% for the top 5\% resource-intensive jobs, slowing training by up to 40\%~\cite{unicorn,cpr}. 
Meta projects MTBF as low as 14 minutes for 130,000-GPU clusters~\cite{reliability-at-scale-meta}, a trend echoed by ByteDance~\cite{megascale}, LAION~\cite{laion}, and Google~\cite{tpu-resiliency}.

\subsection{Checkpointing Techniques}\label{sec:related}

\begin{table}[t!]
    \centering
    \resizebox{\linewidth}{!}{
    \begin{tabular}{@{}l|c|c|c|c@{}}
    \toprule
    \textbf{\makecell{System}} & \textbf{\makecell{Low Overhead \&\\High Frequency}} & \textbf{\makecell{Fast\\Recovery}} & \textbf{\makecell{Full\\Recovery}} & \textbf{\makecell{High\\ETTR}}  \\ \midrule
    CheckFreq~\cite{checkfreq} & \redcross & \redcross & \greencheck & \redcross \\
    Gemini~\cite{gemini} & \redcross & \redcross & \greencheck & \redcross \\
    MoC-System~\cite{moc-system} & \redcross & \greencheck & \redcross & \redcross \\
    \NAME & \greencheck & \greencheck & \greencheck & \greencheck \\ \bottomrule
    \end{tabular}
    }
    \vspace{-0.1in} 
    \caption{Comparison of periodic checkpointing techniques.}
    \label{tbl:technique-comparision}
\end{table}

\noindent{\bf Checkpointing for Fault-Tolerant Training.}  
Periodic checkpointing remains the predominant method for achieving fault tolerance in distributed training~\cite{deepspeed, bloom, megascale, unicorn, microsoft-training-trace, lineage, bytecheckpoint}. However, naive checkpointing introduces significant stalls\footnote{Checkpoint-induced stall occurs when checkpoint I/O exceeds the forward/backward pass, delaying the optimizer step until checkpoint completion.}, severely limiting ETTR, particularly for large models. Recent methods like CheckFreq~\cite{checkfreq} and Gemini~\cite{gemini} mitigate this by overlapping checkpoint operations with computation. CheckFreq introduces a two-phase checkpointing pipeline, dynamically adapting checkpoint intervals based on runtime measurements to balance overhead and recovery costs. Gemini employs in-memory checkpointing, utilizing high-bandwidth CPU memory and strategic placement of checkpoints to speed up recovery. Despite these optimizations, both methods incur substantial runtime overhead when frequently checkpointing large MoE models due to their significantly increased checkpoint sizes, as illustrated in Figure~\ref{fig:checkpoiniting-overheads}. Consequently, neither can fully overlap large MoE checkpoint snapshots with computation, resulting in long intervals and persistently low ETTR—thus failing to adequately address the runtime–recovery tradeoff (Challenge \#1).

More recently, MoC-System~\cite{moc-system} proposed Partial Expert Checkpointing (PEC) for MoE models. PEC reduces checkpoint size by snapshotting only a subset of experts each iteration in a round-robin fashion. While initially effective, PEC compromises correctness during recovery: experts lacking recent checkpoints revert to stale states, causing token loss and breaking synchronous training semantics. Although MoC-System attempts to mitigate accuracy degradation by adaptively increasing the number of experts checkpointed after each failure, this approach rapidly devolves into dense checkpointing under frequent failures, nullifying its initial efficiency advantage. As a result, MoC-System struggles to balance correctness and efficiency (Challenge \#2).

Other optimizations include storage-level enhancements such as NVMe-driven checkpoint acceleration (Fast-Persist~\cite{fast-persist}), concurrent data-parallel checkpoint sharing (Megascale~\cite{megascale}) and persistence (PCCheck~\cite{pccheck}), parallelism-agnostic checkpoint representations (ByteCheckpoint~\cite{bytecheckpoint}), and checkpoint size reduction via model-specific quantization (Check-N-Run~\cite{check-n-run}, CPR~\cite{cpr}). Yet, these techniques still require checkpointing the entire model state simultaneously, inevitably causing stalls and limiting checkpoint frequency.

\noindent{\bf Checkpoint-less Fault-Tolerance.}  
An orthogonal approach to traditional checkpointing exploits redundancy in training setups to minimize or eliminate periodic checkpoints entirely. Just-in-Time~\cite{just-in-time} leverages redundancy across data-parallel replicas, deferring checkpointing until failures are actually detected. More aggressive checkpoint-less strategies include Bamboo~\cite{bamboo}, which adds redundant computations to withstand preemptions in cloud spot instances, Oobleck~\cite{oobleck}, which uses heterogeneous pipelines to recover from failures without requiring spare resources, and ReCycle~\cite{recycle}, which dynamically reroutes workloads to functionally redundant data-parallel peers to continue computation after failures.

However, these checkpoint-less methods critically depend on the presence of redundancy, limiting their applicability as memory optimizations such as Fully Sharded Data Parallelism (FSDP) and ZERO-style optimizations~\cite{deepspeed, zero, zero-infinity, pytorch-fsdp} deliberately eliminate it. This creates a strong need for checkpoint-based techniques in redundancy-constrained training setups.

\begin{figure}[t!]
    \centering
    \includegraphics[width=1.0\linewidth]{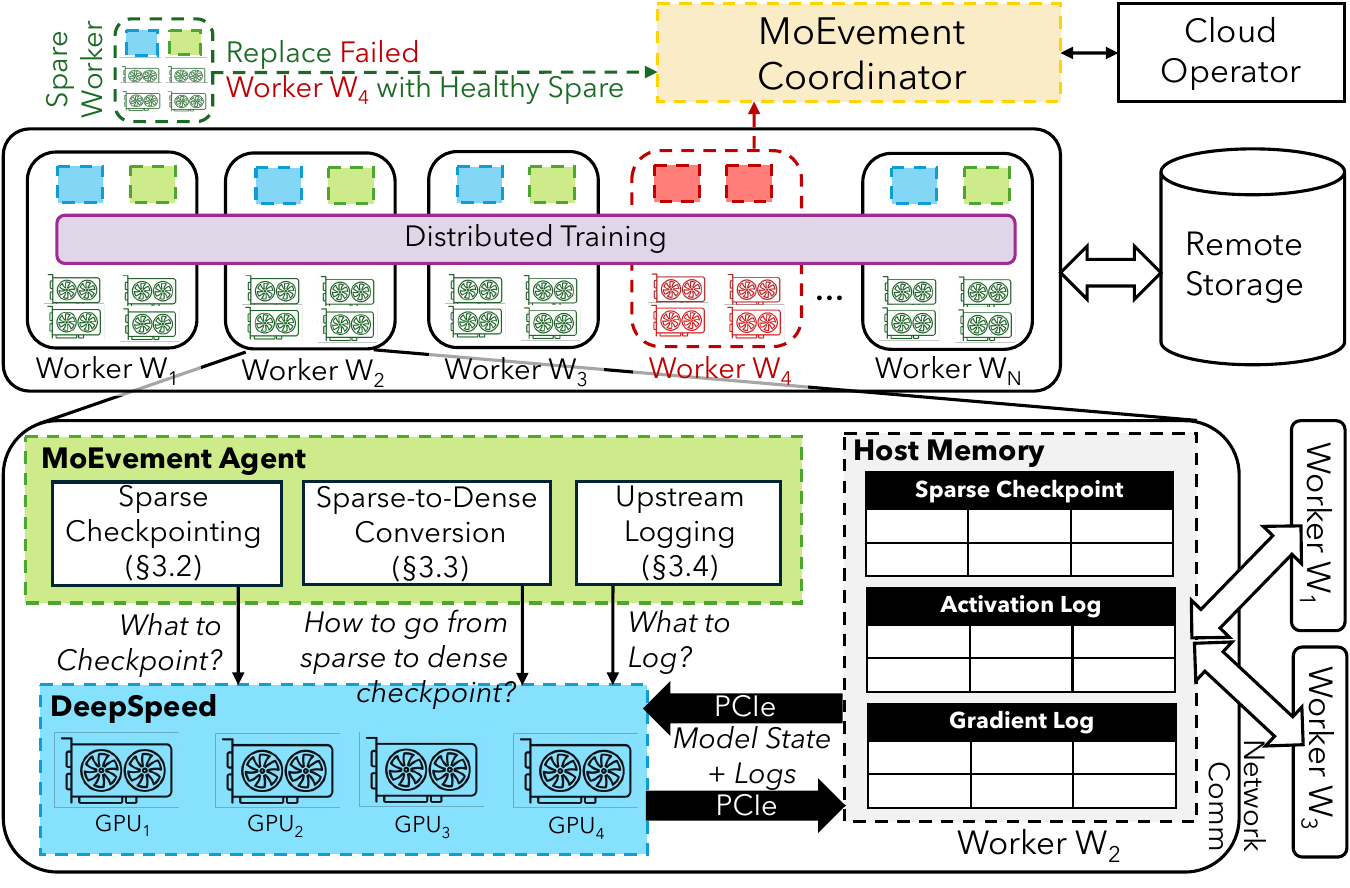}
    \vspace{-0.25in} 
    \caption{The system architecture of \NAMESC}
    \label{fig:overview}
\end{figure}

\subsection{How MTBF Affects ETTR}\label{sec:ettr-mtbf}

Effective Training Time Ratio (ETTR) depends on two factors: runtime overhead from checkpointing during fault-free operation and recovery overhead from failures. Modeling failures as a Poisson process with rate \(\frac{1}{\text{MTBF}}\), ETTR can be approximated by~\cite{fault-tolerance-techniques}:
\[
\mathrm{ETTR}\approx
\frac{1}{1+\underbrace{\frac{T_\text{ckpt}}{T_\text{iter} \times \text{Ckpt}_\text{interval}}}_{\text{Runtime Overhead}}}
\label{eq:ettr-mtbf} \; \times \; \frac{1}{1+\underbrace{\frac{\mathbb{E}[R]}{\text{MTBF}}}_{\text{Recovery Overhead}}}
\]
where \(T_\text{iter}\) is the iteration time, \(\frac{T_\text{ckpt}}{T_\text{iter} \times \text{Ckpt}_\text{interval}}\) is the runtime overhead from checkpointing every \(\text{Ckpt}_{\text{interval}}\) iterations, and \(\mathbb{E}[R]\) is the expected recovery time per failure which scales linearly with checkpoint interval, on-average half the interval (\(\mathbb{E}[R] \approx \tfrac{1}{2} \times {\text{Ckpt}}_{\text{interval}} \times T_{\text{iter}}\))~\cite{daly-2006}. These relationships imply an inherent trade-off in selecting checkpoint intervals: longer intervals reduce runtime overhead but increase expected recomputation after failures, and vice versa. The \emph{optimal checkpoint interval} balances these competing costs to maximize ETTR, thereby achieving faster end-to-end training. As seen in Fig.~\ref{fig:ettr-mtbf}, as MTBF decreases, recovery overhead starts to dominates, pushing ETTR downward and shifting optimal checkpoint intervals toward shorter durations. 
\section{The \NAMESC Approach}\label{sec:technique}

\subsection{\NAMESC Overview}\label{sec:overview}

\NAME is a distributed, in-memory checkpointing system tailored to the 
expert-parallel architecture of MoE models. It reduces checkpointing overhead by exploiting the dynamic and skewed token-to-expert assignment patterns inherent in MoE training, while preserving synchronous training semantics and ensuring fast and accurate failure recovery.

\NAME introduces three key ideas. First, \emph{Sparse Checkpointing} \cref{sec:sparse-checkpointing} incrementally captures subsets of operators across multiple iterations instead of checkpointing the full training state at once. By spreading the work over a sparse checkpointing window, it incurs negligible runtime overhead during fault-free execution, enabling high-frequency, continuous checkpointing. Second, \emph{Sparse-to-Dense Checkpoint Conversion} \cref{sec:sparse-to-dense-conversion} resolves the temporal inconsistency of sparse snapshots by incrementally reconstructing a logically consistent dense checkpoint—selectively replaying computations and activating operators as their master weights and optimizer state become available. This technique ensures correctness and recovery semantics of dense checkpointing without incurring its overhead. Third, \emph{Upstream Logging} \cref{sec:upstream-logging} captures input activations flowing forward and gradients propagating backward at each pipeline-stage boundary. During recovery, these logs enable each stage to independently recompute the state of failed workers without rolling back unaffected ones.

\subsection{Avoiding Stalls with Sparse Snapshots}\label{sec:sparse-checkpointing}

\begin{figure}[t]
    \centering
    \includegraphics[width=1.0\linewidth]{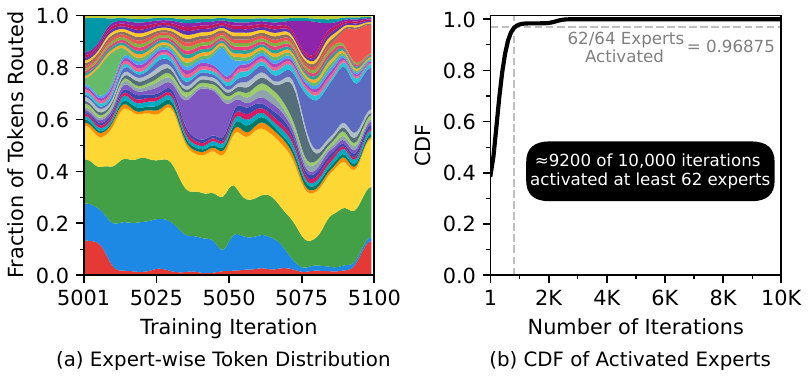}
    \vspace{-0.25in} 
    \caption{MoE training dynamics in DeepSeek-16.4B/64E~\cite{deepseek-moe}. (a) Token distribution (color-coded by expert) is dynamic and skewed. (b) CDF of activated experts shows that nearly all experts are active in most iterations, each receiving non-zero tokens with uneven shares.}
    \label{fig:expert-popularity}
\end{figure}

\begin{figure}[t]
    \centering
    \begin{subfigure}[t]{1.0\linewidth}
        \centering
        \includegraphics[width=\linewidth]{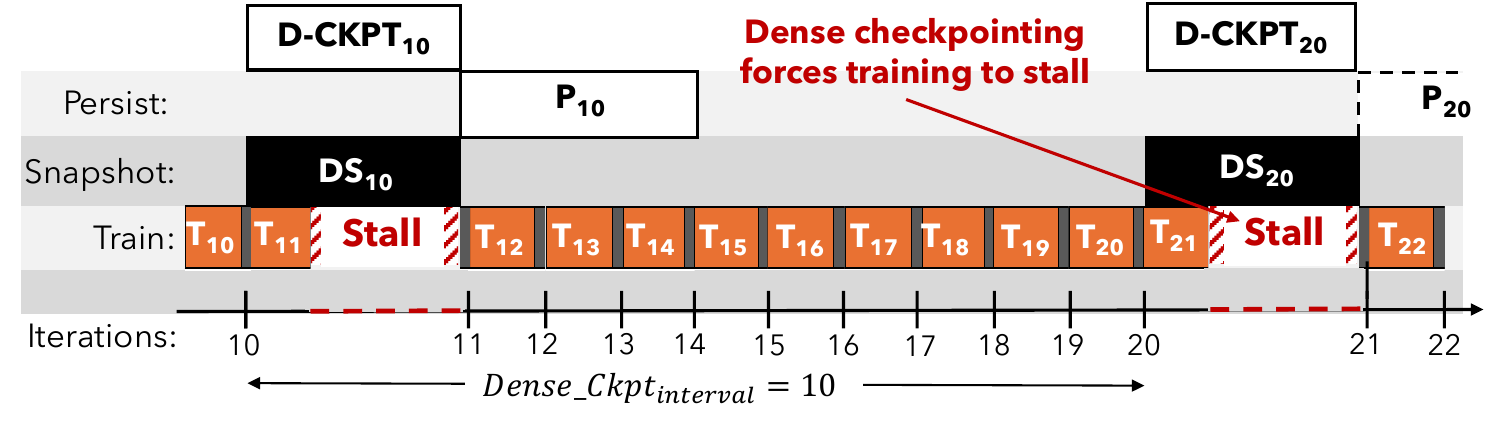}
        \caption{Dense checkpointing stalls training; checkpoints occur infrequently.}
        \label{fig:dense-checkpointing-gantt}
    \end{subfigure} 
    \hfill
    \begin{subfigure}[t]{1.0\linewidth}
        \centering
        \includegraphics[width=\linewidth]{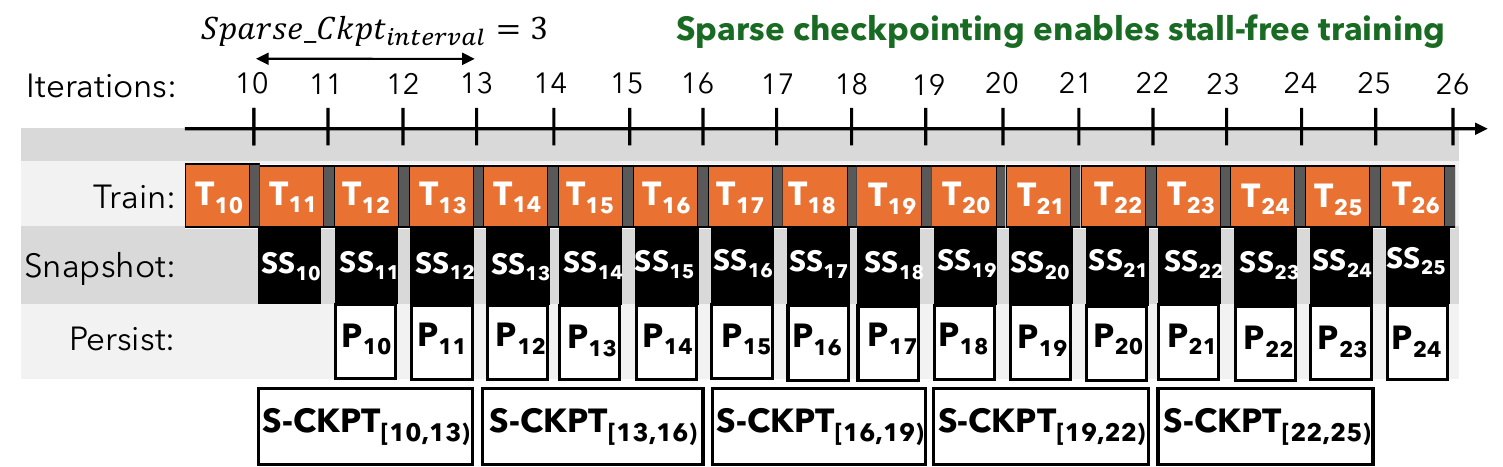}
        \caption{Sparse checkpointing fully overlaps with training, eliminating stalls and enabling frequent checkpoints.}
        \label{fig:sparse-checkpointing-gantt}
    \end{subfigure} 
    \vspace{-0.1in} 
    \caption{Dense vs. Sparse checkpointing}
    \label{fig:checkpointing-gantt}
\end{figure}

\begin{figure}[t!]
    \centering
    \includegraphics[width=1.0\linewidth]{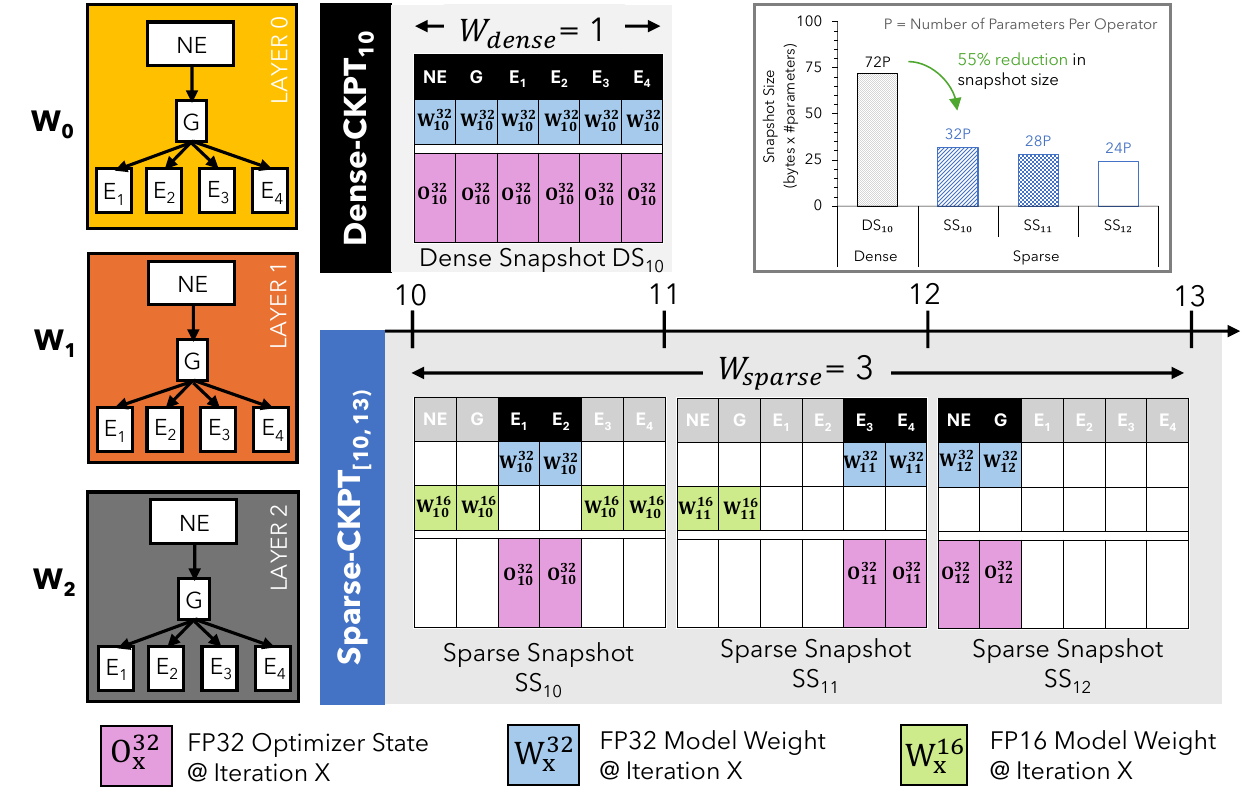}
    \caption{Dense checkpointing snapshots FP32 optimizer state and model weights for all operators in a single iteration. Sparse checkpointing incrementally snapshots FP32 states of different operator subsets over three iterations. \textbf{Inset (top-right):} Sparse checkpointing reduces the per-snapshot size by $55\%$ relative to dense.}
    \label{fig:dense-vs-sparse-snapshot}
\end{figure}

\NAME exploits two key insights to achieve low-overhead, high-frequency checkpointing. First, token distribution across experts is dynamic and skewed: nearly all experts are \emph{active} (assigned at least one token) in most steps ($\geq62/64$ in $\approx$9.2K/10K iterations), yet token shares fluctuate widely (Fig.~\ref{fig:expert-popularity}). \NAME strategically leverages this imbalance, as detailed in Section~\ref{sec:checkpointing-policy}. Second, the iterative nature of training allows reconstructing an operator’s full training state (model weights and optimizer state) from earlier checkpoints by replaying micro batches. Rather than simultaneously checkpointing all operators, \NAME incrementally checkpoints subsets of operators over multiple iterations. At each iteration, only a small subset of operators checkpoint their training state; others checkpoint only their compute weights—83\% smaller (2 bytes vs. 12 bytes per parameter in FP16-FP32 mixed-precision training using Adam optimizer~\cite{adam})—and temporarily enter a \emph{frozen} state\footnote{\emph{Frozen} operators skip weight-gradient computations and optimizer updates, performing only forward and input-gradient computations.}. Transitions into and out of this \emph{frozen} state are orchestrated by the sparse-to-dense conversion mechanism detailed in~\cref{sec:sparse-to-dense-conversion}, cutting per-iteration checkpoint sizes by $\approx55\%$ (Fig.~\hyperref[fig:dense-vs-sparse-snapshot]{\ref*{fig:dense-vs-sparse-snapshot}(Inset)}), fully overlapping checkpoint I/O with computation, and eliminating stalls (Fig.~\ref{fig:sparse-checkpointing-gantt}).

We define a \emph{snapshot} as moving the training states from GPU to local CPU memory during training. Traditional \emph{dense} checkpointing methods snapshot the entire training state, model weights and optimizer state, within a single iteration (\( W_{\text{dense}} = 1 \)). If the snapshot is not completed before the subsequent optimizer step, training stalls, which forces long intervals to amortize cost (\eg \(\text{Ckpt}_\text{interval}=10\) in Fig.~\ref{fig:dense-checkpointing-gantt}).

Sparse checkpointing treats each expert (\(\text{E}_1\)–\(\text{E}_4\)), non-expert (\(\text{NE}\)), and gating (\(\text{G}\)) operator as independently snapshotable. As illustrated for a three‑layer MoE under FP16-FP32 mixed-precision training (Fig.~\ref{fig:dense-vs-sparse-snapshot}), iteration 11: dense \(\mathrm{DS}_{10}\) would snapshot all operators at once, whereas sparse $\text{SS}_{10}$ snapshots $E_1,E_2$'s FP32 states and $E_3,E_4, \text{NE},\text{G}$'s FP16 compute weights; iteration 12 ($\text{SS}_{11}$) snapshots $E_3,E_4$'s FP32 states and FP16 weights for $\text{NE},\text{G}$; iteration 13 ($\text{SS}_{12}$) snapshots $\text{NE},\text{G}$'s FP32 states. After three iterations ($W_{\text{sparse}}=3$), each operator has exactly one FP32 snapshot.

{\bf Persisting Snapshots.} Once a sparse snapshot is moved to local CPU memory, \NAME asynchronously persists it to remote CPU memory on \(r\) peer nodes (similar to Gemini~\cite{gemini}), concurrently with training. A sparse checkpoint is considered persisted once all snapshots within \(W_{\text{sparse}}\) window are durably replicated. \NAME always maintains one persisted checkpoint and another in-flight, garbage-collecting the oldest checkpoint after persisting a new one.

\subsection{Sparse-to-Dense Checkpoint Conversion}\label{sec:sparse-to-dense-conversion}

Sparse checkpointing avoids stalls but introduces temporal inconsistencies: operators are snapshotted at different iterations (\eg in $\text{S-CKPT}_{[10,13)}$ operators $E_1, E_2$ at iteration 10, $E_3, E_4$ at iteration 11, and $\text{NE}, \text{G}$ at iteration 12). If left unresolved, these mismatches can yield incorrect recovery states and degrade model accuracy~\cite{cpr, qiao2019, check-n-run}. \NAME addresses this by incrementally reconstructing a logically consistent dense checkpoint from multiple sparse snapshots. 

\begin{figure}[t!]
    \centering
    \includegraphics[width=1\linewidth]{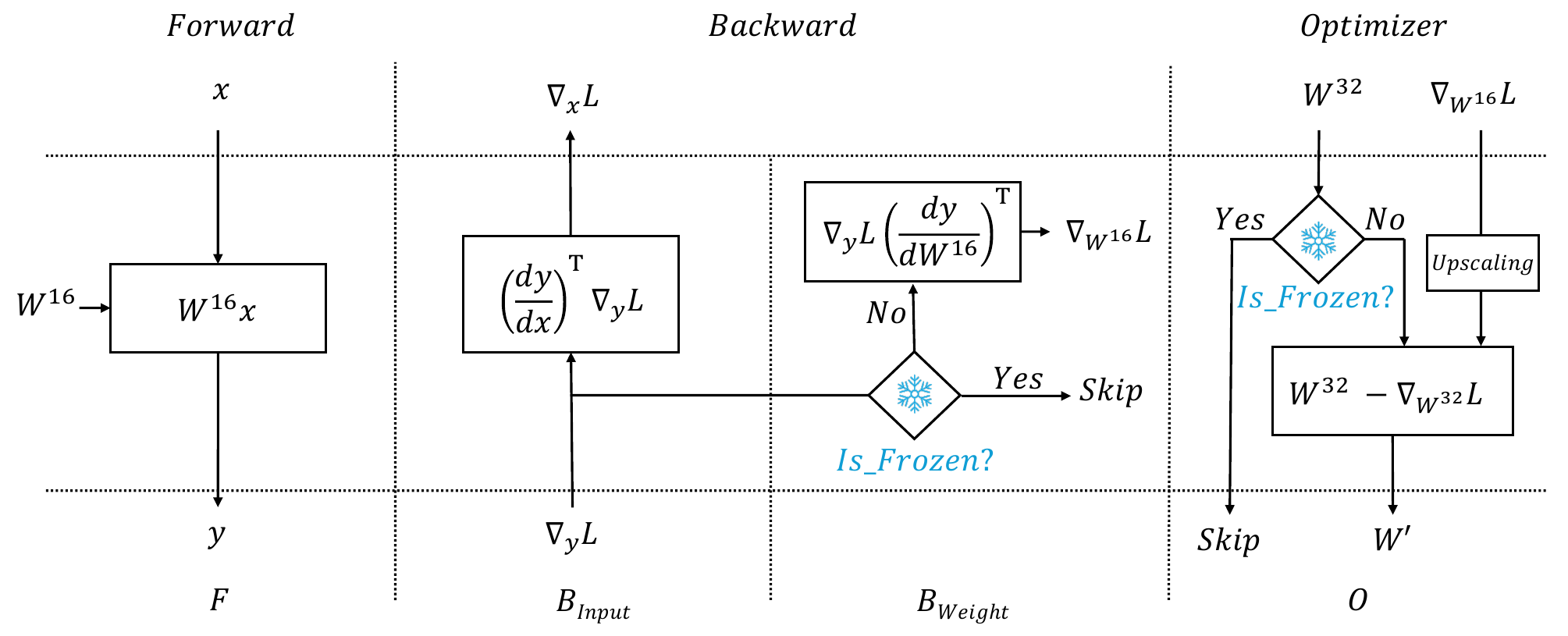}
    \caption{Conditional execution of forward, backward, and optimizer steps based on operator state (\emph{frozen} vs. \emph{active}).}
    \label{fig:frozen-module}
\end{figure}

We assume standard mixed-precision FP16-FP32 training, where FP16 weights are used for forward and backward computation, and FP32 master weights and optimizer state are updated each optimizer step. The key insight is that an operator's state at iteration $t$ depends entirely on its FP32 state at some earlier iteration $s$ ($s < t$) and the parameter updates applied since. Thus, once an \emph{anchor} snapshot is captured at iteration $s$, future FP32 states can be reconstructed by replaying micro batches and re-applying updates. Crucially, replaying does not require FP32 states for operators not yet updating parameters; FP16 weights alone suffice to compute necessary input gradients. In \NAME, operators are classified based on FP32 state availability at snapshot loading. \emph{Active} operators have FP32 weights and optimizer state, performing forward, backward, and optimizer updates. In contrast, \emph{frozen} operators, with only FP16 weights, perform forward and input-gradient computations but skip weight-gradient computations and optimizer updates until a later \emph{anchor} snapshot provides their FP32 state, at which point they become \emph{active} (Fig.~\ref{fig:frozen-module}).

\begin{figure}[t!]
    \centering
    \begin{subfigure}[t]{1.0\linewidth}
        \centering
        \includegraphics[width=\linewidth]{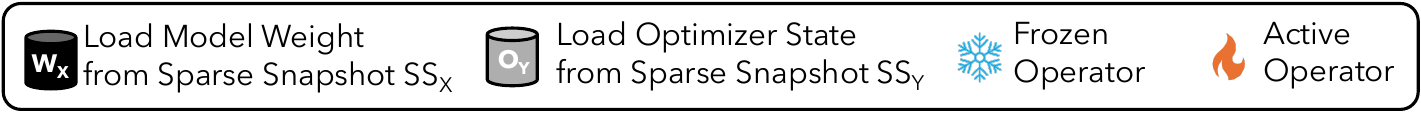}
    \end{subfigure} 
    \vspace{-0.0725in}
    \begin{subfigure}[t]{1.0\linewidth}
        \centering
        \includegraphics[width=\linewidth]{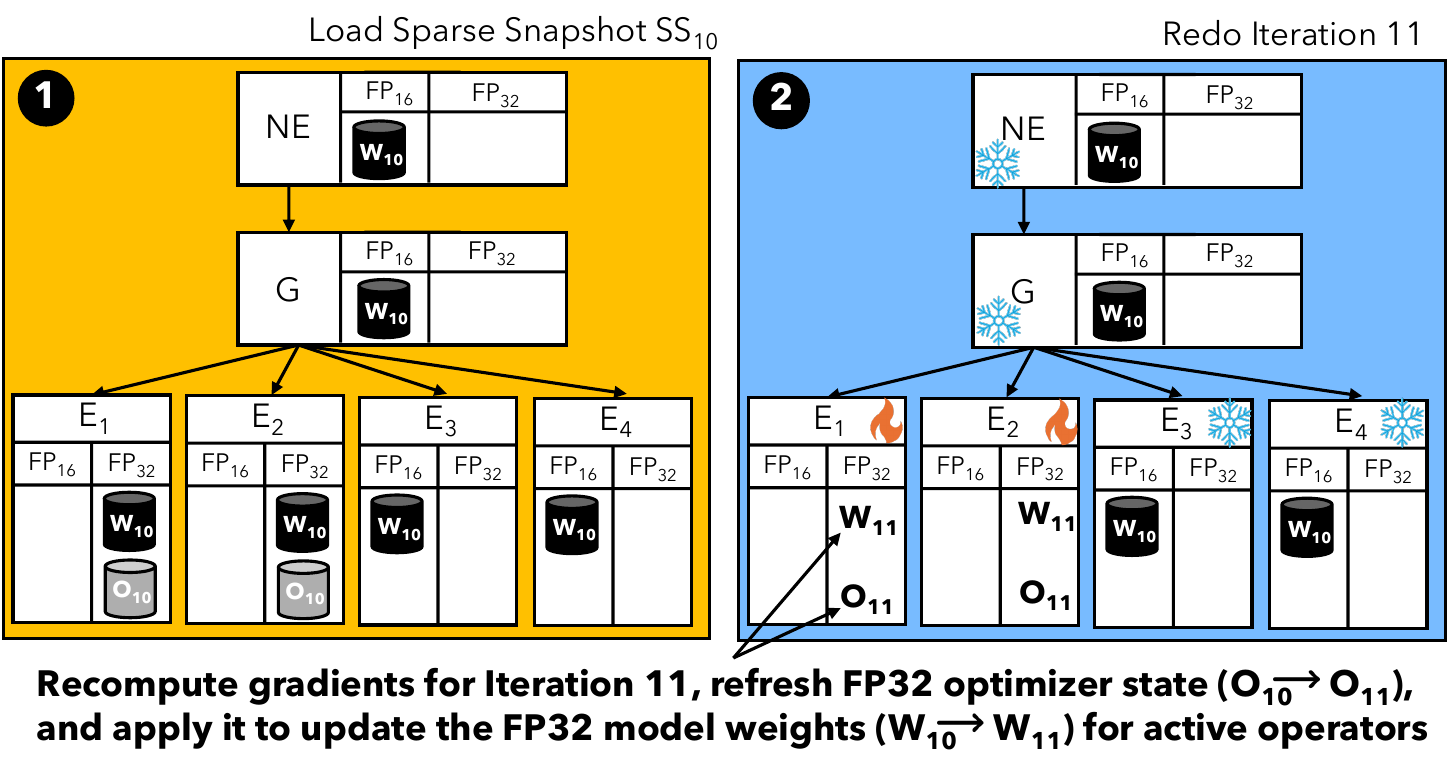}
        \label{fig:sparse-to-dense-1}
        \vspace{-2.0in}
    \end{subfigure} 
    \begin{subfigure}[t]{1.0\linewidth}
        \centering
        \includegraphics[width=\linewidth]{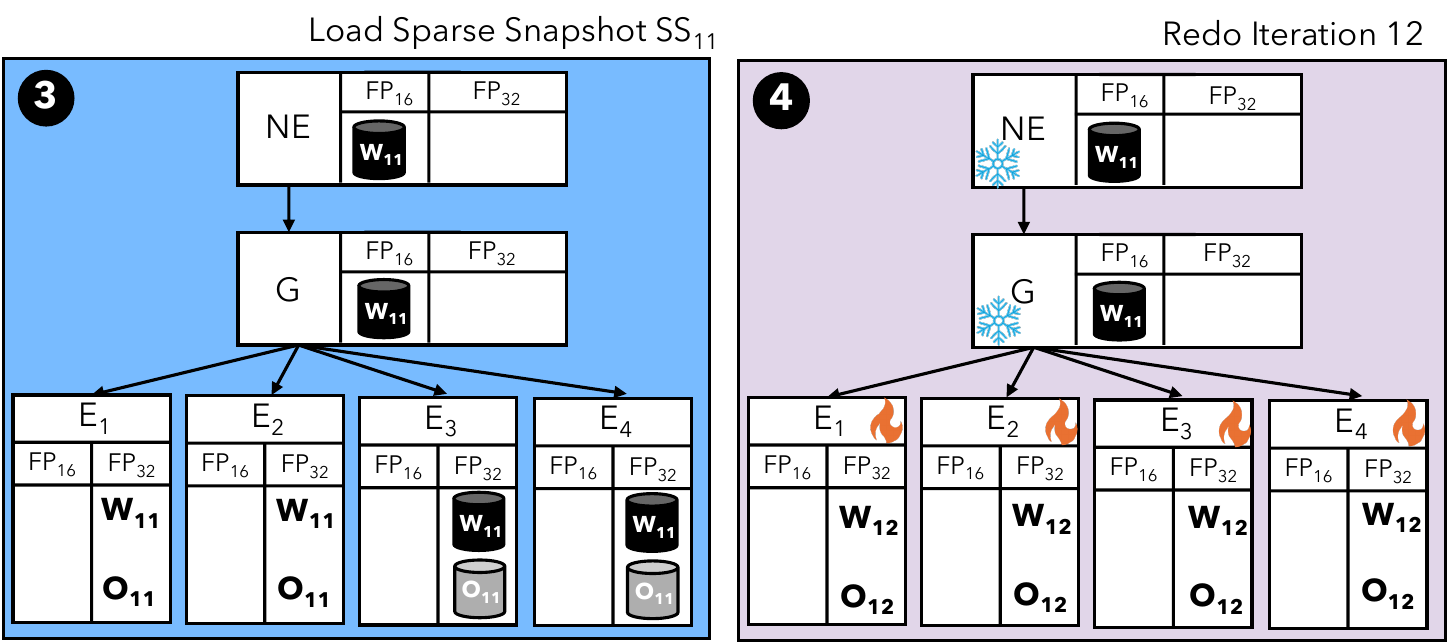}
        \label{fig:sparse-to-dense-2}
        \vspace{-2.0in}
    \end{subfigure} 
    \begin{subfigure}[t]{1.0\linewidth}
        \centering
        \includegraphics[width=\linewidth]{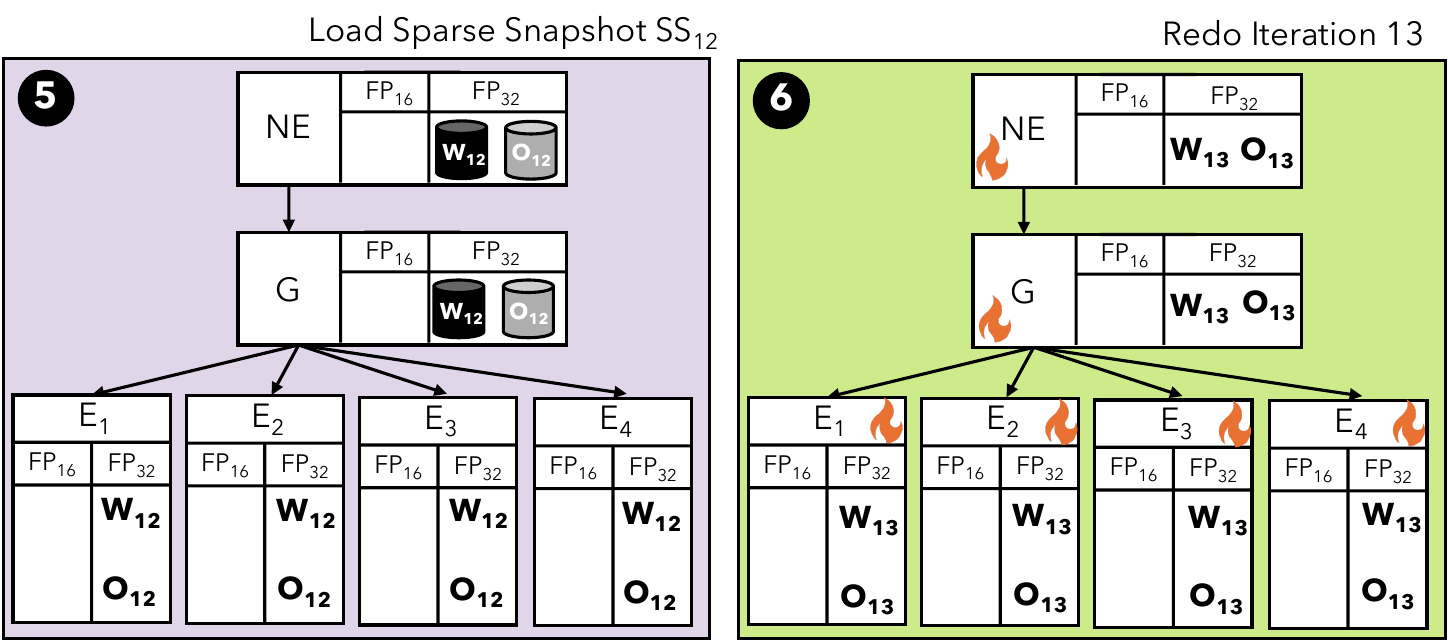}
        \label{fig:sparse-to-dense-3}
        \vspace{-1.0in}
    \end{subfigure} 
    \caption{Step-by-step sparse-to-dense checkpoint conversion over iterations 11–13. Operators transition from \emph{frozen} to \emph{active} as their FP32 model weights and optimizer state become available.}
    \label{fig:sparse-to-dense}
\end{figure}

Figure~\ref{fig:sparse-to-dense} illustrates incremental reconstruction of dense checkpoint at iteration 13 ($\text{D-CKPT}_{13}$) from sparse snapshots ($\text{SS}_{10}$–$\text{SS}_{12}$). Loading snapshot $\text{SS}_{10}$ activates operators $E_1$ and $E_2$ (FP32 state available), leaving operators $E_3$, $E_4$, $G$, and $NE$ frozen with FP16 weights (\numcircledtikz{1}). Replaying micro batches of iteration 11 updates the FP32 weights and optimizer state of active operators ($E_1$, $E_2$), while frozen operators ($E_3$, $E_4$, $NE$,and $G$) participate only in forward and input-gradient computation (\numcircledtikz{2}). Next, loading snapshot $\text{SS}_{11}$ activates operators $E_3$ and $E_4$ alongside $E_1$ and $E_2$, while $G$ and $NE$ remain frozen (\numcircledtikz{3}). Replaying micro batches of iteration 12 updates parameters and optimizer state for active operators ($E_1$–$E_4$), advancing them to iteration 12, whereas frozen operators ($NE$ and $G$) continue propagating gradients without updates (\numcircledtikz{4}). Finally, loading snapshot $\text{SS}_{12}$ activates the remaining operators $G$ and $NE$, making all operators active (\numcircledtikz{5}). Replaying micro batches of iteration 13 updates all operators, achieving a consistent dense checkpoint identical to one captured by traditional dense checkpoint technique (\numcircledtikz{6}).

\subsection{Upstream Logging for Localized Recovery}\label{sec:upstream-logging}

\begin{figure}[t!]
    \centering
    \begin{subfigure}[t]{1.0\linewidth}
        \includegraphics[width=1.0\linewidth]{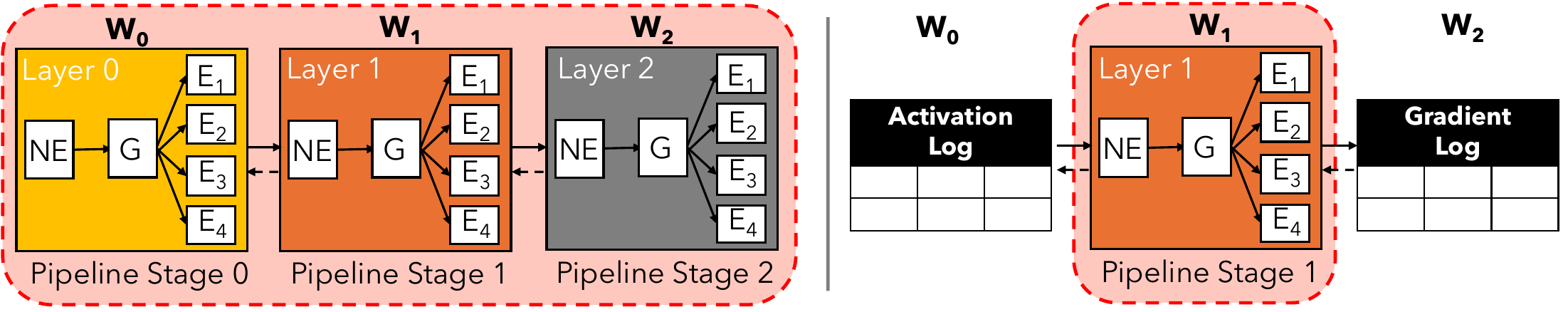}
        \caption{Recomputation scope \textit{without} (left) and \textit{with} (right) Upstream Logging} 
        \label{fig:recovery-scope}
    \end{subfigure}
    \hfill
    \begin{subfigure}[t]{1.0\linewidth}
        \includegraphics[width=1.0\linewidth]{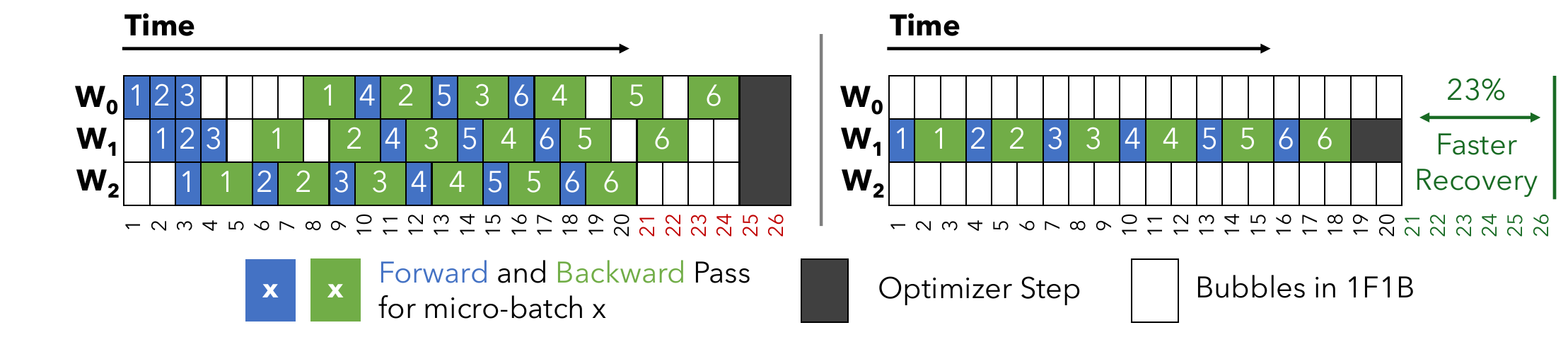}
        \caption{1F1B schedule \textit{without} (left) and \textit{with} (right) Upstream Logging} 
        \label{fig:1f1b-schedule}
    \end{subfigure}
    \vspace{-0.1in} 
    \caption{Upstream Logging narrows recomputation scope to the affected worker (\(W_1\)), achieving \(23\%\) faster recovery.}
    \label{fig:upstream-logging}
\end{figure}

Fast recovery from failures is essential to sustaining high training efficiency. In dense checkpointing, a failure at any pipeline stage triggers rollback of all stages, forcing even non-faulty workers to redo computations~\cite{checkfreq, gemini, megascale}. This cascaded rollback inflates recomputation and prolongs recovery times, as 1F1B schedules introduce pipeline bubbles while re-priming the pipeline (Figure~\ref{fig:1f1b-schedule}, left). For example, recovering stage $W_1$ in a three-stage pipeline requires rolling back $W_0$, $W_1$, and $W_2$ (Figure~\ref{fig:recovery-scope}, left), a penalty that grows as pipeline depth increases—a common occurrence as large models scale beyond a single node using pipeline parallelism~\cite{pipedream, megatron-sc}.

\NAME avoids cluster-wide rollback with \emph{Upstream Logging}, a lightweight mechanism that restricts rollback to only the affected data-parallel groups. During training, \NAME logs intermediate tensors at each pipeline stage boundary: (1) activations passed downstream during forward propagation, and (2) gradients sent upstream during backward propagation. Logs are recorded at the sender stage and stored off GPU in host (CPU) memory. Because logging occurs locally at the sender, it introduces no additional network communication and remain accessible even if upstream or downstream workers fail. Each activation and gradient log is tagged with iteration numbers and micro batch identifiers to allow precise and ordered replay during localized recovery.

When a failure occurs, \NAME pauses all data-parallel (DP) groups, aborts the current iteration, and replaces the failed node with a spare. Only the affected DP group rolls back to its most recent sparse checkpoint—usually just a few iterations old—while other groups remain paused in a consistent state. The failed stage then replays computations locally using logs stored on its upstream and downstream neighbors, reconstructing a consistent dense checkpoint without requiring global recomputation. As shown in Figure~\ref{fig:recovery-scope} (right), when $W_1$ fails, recomputation stays confined to $W_1$, reducing recovery latency by \(23\%\) (Figure~\ref{fig:1f1b-schedule}, right), by avoiding pipeline bubbles during the start-up and cool-down phases. Similarly, when multiple concurrent failures occur, affected DP groups independently perform localized recovery in parallel, without global coordination (additional details in Appendix~\ref{appendix:concurrent-failures}).

\paragraph{Stale Log Cleanup.} Logged tensors from prior sparse checkpoints become obsolete once a new sparse checkpoint is persisted, typically every $W_{\text{sparse}}$ iterations. To avoid unbounded accumulation, \NAME proactively garbage collects stale logs from host (CPU) memory. For the MoE models evaluated in \S\ref{sec:technique-breakdown}, these logged tensors occupy less than $2\%$ of available host (CPU) memory (see Table~\ref{tbl:memory-overhead}).

\subsection{Sparse Checkpointing Policy}\label{sec:checkpointing-policy}

To implement sparse checkpointing effectively, \NAME jointly optimizes two key parameters: the checkpoint window size (\(W_{\text{sparse}}\)) and the order in which operators are checkpointed. Together, these choices determine the checkpointing overhead and the recovery overhead.

\paragraph{Determining \(W_{\text{sparse}}\).}  
In \NAME, the training state is captured incrementally over \(W_{\text{sparse}}\) iterations using {\it sparse checkpointing}, which snapshots only a subset of operators in each iteration. To maintain low-overhead checkpointing, \NAME selects the smallest \(W_{\text{sparse}}\) whose snapshot fits within an iteration, determined by the \texttt{FindWindowSize()} function in Algorithm~\ref{sparse-window-algo}. The function starts with all operators marked as \emph{active} for checkpointing and gradually transitions some to \emph{frozen}, recalculating the snapshot size at each step. For \emph{active} operators, it records FP32 master weights and optimizer state; for \emph{frozen} operators, it stores FP16 model weights. The process terminates once the estimated snapshot time fits within the iteration time, ensuring checkpointing does not force training to stall.

\begin{codealgorithm}[t]
\vspace{-0.1in}
\caption{Sparse Checkpoint Scheduling}\label{sparse-window-algo}
\begin{lstlisting}[style=pseudocode]
# O: List of operators in MoE model
# T_Iter: Iteration Time
# S_Compute: Size of compute weights per operator used in Forward/Backward Pass
# S_Master: Size of master weights per operator
# S_Optim: Size of optimizer state per operator
# B_PCIe: Effective GPU-to-CPU PCIe bandwidth
def FindWindowSize(O):
    O_Total, T_Iter, S_Compute, S_Master, S_Optim, B_PCIe = Profiler(O) #get profiled stats
    O_Active = O_Total #all operators active
    while O_Active > 2:
        O_Frozen = O_Total - O_Active
        ckpt_size = (S_Master + S_Optim) * O_Active 
                    + S_Compute * O_Frozen
        if ckpt_size / B_PCIe <= T_Iter:
            break #checkpoint fits within iter time
        O_Active -= 1 #try fewer active operators
    W_Sparse = ceil(O_Total / O_Active)
    return W_Sparse, O_Active
def GenerateSchedule(O, W_Sparse, O_Active):
    O_Ordered = OrderOperators(O) #popularity sort
    schedule = []
    for i in range(W_Sparse):
        start = i * O_Active
        end = min(start + O_Active, len(O))
        schedule.append({
            'active': O_Ordered[start:end],
            'frozen': O_Ordered[end:]
        })
    return schedule
def SparseCheckpointSchedule(O):
    W_Sparse, O_Active = FindWindowSize(O)
    return GenerateSchedule(O, W_Sparse, O_Active)
\end{lstlisting}
\vspace{-0.1in} 
\label{alg:scheduler}
\end{codealgorithm}

\paragraph{Determining Operator Ordering.}  
Expert popularity is measured by the frequency with which an expert is activated.  For expert \(j\) in layer \(l\), the activation count is:
\vspace{-0.05in}
\[
\mathcal{A}_j^l = \sum_{x_i\in\mathcal{D}}\mathbf{1}\left[\gamma(e_j^l, x_i)=1\right]
\]
where \(\mathcal{D}\) is the training dataset and \(\gamma(e_j^l, x_i)\) indicates whether expert \(j\) in layer \(l\) is activated for token \(x_i\). Popularity distributions are naturally skewed—some experts are activated more frequently than others, and are commonly referred to as \emph{popular} experts~\cite{flexmoe, fastmoe}. \NAME opportunistically exploits this imbalance by sorting experts in ascending order of popularity using the \texttt{OrderOperators()} function, deferring the checkpointing of popular experts to later iterations within the sparse checkpointing window. Deferring popular experts keeps them \emph{frozen} longer during sparse-to-dense conversion, avoiding weight-gradient and optimizer updates and lowering recomputation cost by $\approx33\%$ compared to \emph{active} experts. Alternative ordering strategies are described in Appendix~\ref{appendix:ordering-scheme}.


Expert popularity can evolve over the course of training~\cite{smartmoe, fastermoe}. \NAME reorders operators when activation frequencies change by over $10\%$ for at least $25\%$ of experts, balancing responsiveness with schedule stability (evaluated in \S\ref{sec:technique-breakdown}). We further analyze the impact of expert popularity skewness on \NAME's performance in Appendix~\ref{appendix:popularity-effect}, and discuss generalizing our checkpointing policy to dense models in Appendix~\ref{appendix:generalizing-to-dense}.

\paragraph{Runtime Analysis.} The overall complexity of Algorithm~\ref{sparse-window-algo} is $\bigO(|O| \log |O|)$, where $|O|$ is the number of operators. The \texttt{FindWindowSize()} functions runs in $\bigO(|O|)$ time, with the dominant cost coming from sorting operators by popularity during schedule generation. In practice, the algorithm runs on the CPU, completing in (\(\approx0.1\) sec), and executes asynchronously without interrupting GPU computation.

\subsection{Recovery Guarantees under Sparse Checkpointing}
When a failure occurs, training resumes from the most recent checkpoint. In existing techniques, a dense checkpoint is taken every ${\text{Ckpt}}_{\text{interval}}$ iterations. Assuming failures occur uniformly at random, let $R$ denote the recovery time following a failure and $T_{\text{iter}}$ the duration of a single iteration. For existing checkpointing techniques, $R$ is bounded by the checkpoint interval, and the expected recovery time $\mathbb{E}[R]$ is, on-average, half the checkpoint interval~\cite{daly-2006, checkfreq}:
\[
0 \le R \le {\text{Ckpt}}_{\text{interval}} \times T_{\text{iter}}
\quad \text{and} \quad 
\mathbb{E}[R] \approx \tfrac{1}{2} \times {\text{Ckpt}}_{\text{interval}} \times T_{\text{iter}}
\]
In contrast, \NAME recovers in two phases: (1) replaying $W_{\text{sparse}}$ iterations to reconstruct a dense checkpoint from the most recent sparse checkpoint, and (2) re-executing up to $W_{\text{sparse}}$ additional iterations from that restored state. This yields the following bounds \(R\) and \(\mathbb{E}[R]\):
\[
0 \le R \le 2 \times W_{\text{sparse}} \times T_{\text{iter}}
\quad \text{and} \quad 
\mathbb{E}[R] \approx \tfrac{3}{2} \times W_{\text{sparse}} \times T_{\text{iter}}
\]
Empirically, we find $W_{\text{sparse}} \ll {\text{Ckpt}}_{\text{interval}}$, and demonstrate in \S\ref{sec:training-efficiency} that \NAME checkpoints up to $26\times$ more often than dense checkpointing techniques like Gemini and CheckFreq.
\section{Implementation}\label{sec:implementation}

We implement \NAME on top of DeepSpeed~v0.16~\cite{deepspeed}, adding $\approx2K$ lines of Python code.

\paragraph{Sparse Snapshot.} We modify DeepSpeed's checkpoint saving routine to operate at per-operator granularity, treating each expert, non-expert, and gating operator as independently snapshotable. For \emph{active} operators, we record master weights and optimizer states; for \emph{frozen} operators, only compute weights. GPU-to-CPU transfers use pinned host buffers, allocated once at the beginning of the training job and reused throughout, with asynchronous \texttt{cudaMemcpyAsync} calls on a dedicated CUDA stream to overlap device-to-host I/O with forward and backward passes. Once in host memory, snapshots are replicated to $r$ peer nodes (default $r=2$) via \texttt{torch.distributed} point-to-point (\texttt{send}/\texttt{recv}), with replication running asynchronously alongside training.

\paragraph{Sparse-to-Dense Conversion.} We integrate sparse-to-dense conversion into DeepSpeed's checkpoint loading routine. On load, each operator is marked \emph{active} if their master weights and optimizer states are available, otherwise marked \emph{frozen}. \emph{Active} operators run forward, backward, and optimizer update steps, while \emph{frozen} operators perform only forward and input-gradient computations. We extend DeepSpeed with state-aware execution paths, wrapping gradient computation and optimizer updates in state checks, using \texttt{torch.no\_grad()} to skip autograd tracking for \emph{frozen} operators. During recovery, the replay engine loads sparse snapshots in schedule order, replays the corresponding micro batches, and transitions operators to \emph{active} state once their full state is restored from the sparse snapshot. This continues until all operators are marked \emph{active}, yielding a dense checkpoint.

\paragraph{Upstream Logging.} We extend DeepSpeed's \texttt{PipelineModule} to log a copy of activations and gradients in pinned host buffers, tagged with iteration and micro batch ID. DeepSpeed retains these tensors in GPU memory while transmitting them to the next pipeline stage; we leverage this period to issue an asynchronous \texttt{cudaMemcpyAsync} on a dedicated CUDA stream, overlapping GPU-to-CPU transfers with ongoing training. Logged tensors remain in host memory until their corresponding sparse checkpoint is either consumed during recovery or garbage-collected.
\section{Evaluation}\label{sec:evaluation}

\subsection{Experimental Setup}\label{sec:experimental-setup}

\paraf{Cluster.} Unless otherwise noted, all of our experiments were conducted on a GPU cluster with 12 Standard\_NC96ads\_A100\_v4 nodes, each of which has a 64-core AMD EPYC 7V13 CPU, 880 GB of RAM, and eight Nvidia A100 80 GB GPUs on Azure Cloud. GPUs on the same node are connected via a 600 GB/s NVLink interconnect, and nodes are connected via 80 Gbps inter-node interconnect across 8 NICs. Azure Blob Storage is used as the remote persistent storage and the aggregated bandwidth to it is 40Gbps. All servers run 64-bit Ubuntu 22.04 with CUDA library v12.8.

\paraf{Baselines.} We compare \NAME against Gemini~\cite{gemini}, a state-of-the-art in-memory checkpointing system, CheckFreq~\cite{checkfreq}, a disk-based checkpointing alternative, and MoC-System~\cite{moc-system}, a recent MoE-specific checkpointing technique. To assess potential interference from these checkpointing systems, we also measured the throughput of DeepSpeed-MoE~\cite{deepspeed} with checkpointing disabled, utilizing a 1F1B interleaved schedule and ZERO Stage-1 optimizations~\cite{zero}.

\begin{table}[t!]
    \centering
    \resizebox{\linewidth}{!}{
    \begin{tabular}{@{}c|c|c|c|c|c|c@{}}
    \toprule
    \textbf{Model}
    & \textbf{\#Layers} & \textbf{Gate}
    & \textbf{\makecell{\#Experts\\Per Layer}}
    & \textbf{\makecell{\#Activated\\Per Token}}
    & \textbf{\makecell{Total\\Params}}
    & \textbf{\makecell{Active\\Params}} \\ \midrule
    MoE-LLaVa~\cite{moe-llava}  & 32 & Top-2 & 4 & 2 & 2.9B & 2B \\ 
    GPT-MoE~\cite{deepspeedmoe} & 12 & Top-6 & 32 & 6 & 7.3B & 1.6B \\ 
    QWen-MoE~\cite{qwen2} & 24 & Top-8 & 64 & 8 & 14.3B & 2.7B \\ 
    DeepSeek-MoE~\cite{deepseek-moe} & 28 & Top-8 & 64 & 2(shared) + 8 & 16.4B & 3.7B \\ \bottomrule
    \end{tabular}
    }
    \caption{Specifications of models used for evaluation.}
    \label{tbl:model-spec}
\end{table}

\paraf{Models.} We evaluate all systems using the four representative MoE models shown in Table~\ref{tbl:model-spec}: MoE-LLaVa~\cite{moe-llava}, GPT-MoE~\cite{gpt-4}, QWen-MoE~\cite{qwen2}, and DeepSeek-MoE~\cite{deepseek-moe}. GPT-MoE, QWen-MoE, and DeepSeek-MoE are trained on RedPajama dataset~\cite{redpajama} with a batch size of 512, micro-batch size of 32, and sequence length of 2048 tokens using parallelization strategies (PP, DP, EP) degrees of (3, 4, 8), (6, 2, 8), and (12, 1, 8) respectively. MoE-LLaVa is trained on the ImageNet-1K dataset~\cite{imagenet-1k}  using (PP, DP, EP) degrees of (6, 2, 8).

\begin{table*}[t!]
  \centering
  \resizebox{\textwidth}{!}{%
  \begin{tabular}{c|cccc|cccc|c|cccc|cccc}
  \toprule
  \multirow{3}{*}{\textbf{Model}} 
  & \multicolumn{4}{c|}{\textbf{Checkpointing Interval}} 
  & \multicolumn{4}{c|}{\textbf{Avg. Per-Iteration Checkpointing Overhead}} 
  & \multirow{3}{*}{\textbf{MTBF}} 
  & \multicolumn{4}{c|}{\textbf{Total Recovery Time}} 
  & \multicolumn{4}{c}{\textbf{Effective Training Time Ratio (ETTR)}} \\ 
  
  & \multicolumn{4}{c|}{\textbf{(iterations)}} 
  & \multicolumn{4}{c|}{\textbf{(seconds, overhead \% in parentheses)}} 
  &  
  & \multicolumn{4}{c|}{\textbf{(seconds)}} 
  & \multicolumn{4}{c}{
    \raisebox{-0.65\height}{
    \begin{tikzpicture}
    \begin{axis}[
        hide axis,
        scale only axis,
        height=0.2cm,
        width=3.5cm,
        colorbar horizontal,
        point meta min=0.0,
        point meta max=1.0,
        colorbar style={
            width=3.5cm,
            xtick={0.0, 0.25, 0.5, 0.75, 1.0},
            xticklabel style={font=\small},
            height=0.2cm,
        },
        colormap={redyellowgreen}{
            rgb(0pt)=(1,0,0)           
            rgb(83.33pt)=(1,1,0)       
            rgb(100pt)=(0,1,0)         
        },
        colorbar sampled,
    ]
    \addplot [draw=none] coordinates {(0,0)};
    \end{axis}
    \end{tikzpicture}
    }
  } \\ \cline{2-9} \cline{11-18}
   & \textbf{CheckFreq} & \textbf{Gemini} & \textbf{MoC} & \textbf{\NAME} 
   & \textbf{CheckFreq} & \textbf{Gemini} & \textbf{MoC} & \textbf{\NAME} & 
   & \textbf{CheckFreq} & \textbf{Gemini} & \textbf{MoC} & \textbf{\NAME} 
   & \textbf{CheckFreq} & \textbf{Gemini} & \textbf{MoC} & \textbf{\NAME} \\ \hline
   \multirow{5}{*}{\textbf{MoE-LLaVa}~\cite{moe-llava}}
   & \multirow{5}{*}{\makecell{57}} & 46 & \multirow{5}{*}{\makecell{1}} & \multirow{5}{*}{\makecell{1\\ ($W_{sparse}$= 3)}} 
   & \multirow{5}{*}{0.03 (2\%)} & 0.02 (1\%) & 0.14 (8\%) & \multirow{5}{*}{0.01 (1\%)} 
   & 2H 
        & 1265 & 938 & 10 & 11 
        & \ApplyGradient{0.945} & \ApplyGradient{0.959} 
        & \ApplyGradient{0.922} & \ApplyGradient{0.981} \\
    & & 38 & & & & 0.07 (4\%) & 2.14 (124\%) & & 1H 
        & 2530 & 1551 & 20 & 24 
        & \ApplyGradient{0.914} & \ApplyGradient{0.939} 
        & \ApplyGradient{0.375} & \ApplyGradient{0.979} \\
    & & 27 & & & & 0.13 (7\%) & 3.41 (201\%) & & 30M 
        & 5059 & 2203 & 41 & 57 
        & \ApplyGradient{0.853} & \ApplyGradient{0.897} 
        & \ApplyGradient{0.328} & \ApplyGradient{0.975} \\
    & & 21 & & & & 0.21 (11\%) & 5.39 (318\%) & & 20M 
        & 7589 & 2570 & 61 & 74 
        & \ApplyGradient{0.792} & \ApplyGradient{0.852} 
        & \ApplyGradient{0.231} & \ApplyGradient{0.971} \\
    & & 17 & & & & 0.31 (15\%) & 5.88 (347\%) & & 10M 
        & 11178 & 4161 & 98 & 117 
        & \ApplyGradient{0.712} & \ApplyGradient{0.764} 
        & \ApplyGradient{0.184} & \ApplyGradient{0.964} \\ \hline

    \multirow{5}{*}{\textbf{GPT-MoE}~\cite{deepspeedmoe}}
   & \multirow{5}{*}{\makecell{78}} & 64 & \multirow{5}{*}{\makecell{1}} & \multirow{5}{*}{\makecell{1\\ ($W_{sparse}$= 3)}} 
   & \multirow{5}{*}{0.03 (1\%)} & 0.03  (1\%) & 0.18 (9\%) & \multirow{5}{*}{0.03 (1\%)}
    & 2H 
        & 1424 & 1121 & 13 & 18 
        & \ApplyGradient{0.937} & \ApplyGradient{0.947}
        & \ApplyGradient{0.919} & \ApplyGradient{0.979} \\
    & & 49 & & & & 0.06 (2\%) & 3.32 (158\%) & & 1H 
        & 2848 & 1718 & 25 & 28 
        & \ApplyGradient{0.903} & \ApplyGradient{0.923} 
        & \ApplyGradient{0.384} & \ApplyGradient{0.976} \\
    & & 36 & & & & 0.15 (5\%) & 4.54 (216\%) & & 30M 
        & 4486 & 2160 & 50 & 78 
        & \ApplyGradient{0.834} & \ApplyGradient{0.868} 
        & \ApplyGradient{0.326} & \ApplyGradient{0.971} \\
    & & 31 & & & & 0.22 (7\%) & 7.90 (376\%) & & 20M 
        & 8543 & 3259 & 76 & 105 
        & \ApplyGradient{0.765} & \ApplyGradient{0.815} 
        & \ApplyGradient{0.220} & \ApplyGradient{0.966} \\
    & & 22 & & & & 0.33 (11\%) & 8.36 (397\%) & & 10M 
        & 13086 & 4627 & 132 & 161 
        & \ApplyGradient{0.687} & \ApplyGradient{0.722} 
        & \ApplyGradient{0.179} & \ApplyGradient{0.959} \\ \hline

    \multirow{5}{*}{\textbf{QWen-MoE}~\cite{qwen2}}
   & \multirow{5}{*}{\makecell{113}} & 89 & \multirow{5}{*}{\makecell{1}} & \multirow{5}{*}{\makecell{1\\ ($W_{sparse}$= 5)}} 
   & \multirow{5}{*}{0.05 (2\%)} & 0.04 (2\%) & 0.19 (9\%) & \multirow{5}{*}{0.04  (2\%)} 
    & 2H 
        & 1170 & 961 & 15 & 17 
        & \ApplyGradient{0.927} & \ApplyGradient{0.934} 
        & \ApplyGradient{0.927} & \ApplyGradient{0.981} \\
    & & 65 & & & & 0.05 (3\%) & 4.25 (170\%) & & 1H 
        & 2340 & 1402 & 30 & 33 
        & \ApplyGradient{0.898} & \ApplyGradient{0.915}
        & \ApplyGradient{0.380} &\ApplyGradient{0.977} \\
    & & 48 & & & & 0.11 (6\%) & 7.10 (284\%) & & 30M 
        & 4680 & 2114 & 60 & 83 
        & \ApplyGradient{0.841} & \ApplyGradient{0.868} 
        & \ApplyGradient{0.325} & \ApplyGradient{0.970} \\
    & & 36 & & & & 0.19 (9\%) & 9.23 (369\%) & & 20M 
        & 7020 & 2331 & 90 & 140 
        & \ApplyGradient{0.783} & \ApplyGradient{0.819} 
        & \ApplyGradient{0.205} & \ApplyGradient{0.963} \\
    & & 27 & & & & 0.25 (13\%) & 9.68 (386\%) & & 10M 
        & 10040 & 3495 & 146 & 207 
        & \ApplyGradient{0.703} & \ApplyGradient{0.717} 
        & \ApplyGradient{0.162} & \ApplyGradient{0.957} \\ \hline
        
    \multirow{5}{*}{\textbf{DeepSeek-MoE}~\cite{deepseek-moe}}
   & \multirow{5}{*}{\makecell{124}} & 92 & \multirow{5}{*}{\makecell{1}} & \multirow{5}{*}{\makecell{1 \\ ($W_{sparse}$= 6)}} 
   & \multirow{5}{*}{0.08  (3\%)} & 0.07 (2\%) & 0.22 (8\%) & \multirow{5}{*}{0.06  (2\%)} 
     & 2H 
        & 992 & 800 & 17 & 22 
        & \ApplyGradient{0.902} & \ApplyGradient{0.910} 
        & \ApplyGradient{0.933} & \ApplyGradient{0.975} \\
    & & 70 & & & & 0.09 (3\%) & 5.75 (198\%) & & 1H 
        & 1984 & 1218 & 35 & 43 
        & \ApplyGradient{0.877} & \ApplyGradient{0.894} 
        & \ApplyGradient{0.349} & \ApplyGradient{0.970} \\
    & & 54 & & & & 0.19 (6\%) & 8.58 (293\%) & & 30M 
        & 3967 & 1879 & 70 & 104 
        & \ApplyGradient{0.827} & \ApplyGradient{0.853} 
        & \ApplyGradient{0.324} & \ApplyGradient{0.965} \\
    & & 38 & & & & 0.25 (8\%) & 12.87 (444\%) & & 20M 
        & 5951 & 1983 & 104 & 167 
        & \ApplyGradient{0.777} & \ApplyGradient{0.814} 
        & \ApplyGradient{0.168} & \ApplyGradient{0.956} \\
    & & 31 & & & &  0.38 (11\%) & 13.65 (470\%) & & 10M 
        & 9402 & 3236 & 176 & 241 
        & \ApplyGradient{0.672} & \ApplyGradient{0.728} 
        & \ApplyGradient{0.134} & \ApplyGradient{0.951} \\
   \bottomrule
  \end{tabular}%
  }
  \caption{Comparison of checkpointing techniques across representative MoE models under varying Mean Time Between Failures (MTBF).}
  \label{tbl:overall}
\end{table*}

\subsection{Training Efficiency Under Controlled Failures}\label{sec:training-efficiency}

We evaluate \NAME under controlled failures during training of the four MoE models listed in Table~\ref{tbl:model-spec}, using a range of failure rates by varying MTBF from 2 hours down to 10 minutes over 12-hour training runs. Table~\ref{tbl:overall} summarizes our results, addressing three questions:

\textit{\textbf{What checkpoint intervals are achievable, and at what overhead?}}
Checkpoint interval—the number of iterations between successive checkpoints—directly impacts checkpoint overhead and recovery performance, while checkpoint window denotes iterations over which a checkpoint is spread, ensuring every operator is snapshotted at least once. For dense checkpointing methods (CheckFreq and Gemini), this window is always 1. For CheckFreq, we configure its policy module to select intervals that cap runtime overhead at $\le 3\%$, resulting in intervals of 57–124 iterations across evaluated models and overhead of $\le 0.08$\,secs/iter. Gemini, in contrast, employs an oracle policy: intervals are selected offline individually for each MTBF to maximize ETTR. This idealized, hindsight-informed selection provides an upper bound on its achievable performance. At MTBF=2H, Gemini selects shorter intervals (46–92 iterations) with lower overhead ($\le 2\%$) than CheckFreq, benefiting from the use of high-bandwidth CPU memory rather than disk. However, as MTBF decreases, Gemini must shorten intervals further (\eg 17–31 iterations at MTBF=10M) to minimize recomputation, causing per-iteration overhead to increase to 11–15\%.

MoC aggressively checkpoints every iteration by partially snapshotting experts, resulting in an effectively unbounded checkpoint window, risking token loss as some experts remain uncheckpointed indefinitely. MoC maintains modest overhead under infrequent failures, but overhead spikes dramatically as failures become frequent and the token-loss budget is exhausted (\eg $\approx4.7\times$ slowdown at MTBF=10M for DeepSeek-MoE). In contrast, \NAME checkpoints every iteration by incrementally snapshotting subsets of operators over a checkpoint window $W_{\text{sparse}}$, selected by Algorithm~\ref{sparse-window-algo}, ensuring all experts are checkpointed per window, eliminating token loss (\cref{sec:sparse-to-dense-conversion}). This evenly distributes checkpoint operation across multiple iterations, maintaining low overhead ($\le2\%$ slowdown) across all models and failure scenarios.

\textit{\textbf{How quickly can we recover from failures?}}
Total recovery time—the cumulative time spent recovering from all failures—depends on both checkpoint intervals and failure frequency. CheckFreq and Gemini incur extended recovery times (9402–13086 secs for CheckFreq, 3236–4627 secs for Gemini at MTBF=10M) due to global rollbacks and recomputation of half their checkpoint intervals, on average, after each failure. MoC's per-iteration checkpointing strategy limits recomputation and thus total recovery time (98–176 secs at MTBF=10M). However, this aggressive strategy comes at the cost of high runtime overhead—up to 470\% at MTBF=10M. In comparison, \NAME restricts rollbacks exclusively to the affected data-parallel group and limits recomputation to $\lceil \frac{3}{2}\times W_{\text{sparse}} \rceil$ iterations, achieving recovery times of 117–241 secs at MTBF=10M—only marginally higher than MoC due to the sparse-to-dense conversion during recovery. Crucially, \NAME preserves all tokens and synchronous training semantics like CheckFreq and Gemini, yet recovers up to $31\times$ and $18\times$ faster, respectively. 

\textit{\textbf{What training efficiency can be achieved under failures?}} ETTR—the fraction of wall‑clock time spent on useful training computation—encapsulates the combined effects of checkpoint overhead and recovery downtime. CheckFreq and Gemini achieve high ETTR (0.90–0.96) at MTBF=2H due to infrequent failures, but their prolonged recovery periods degrade ETTR under frequent failures. CheckFreq drops to 0.672, Gemini to 0.728 at MTBF=10M. MoC starts strong at low failure rates, but its high checkpoint overhead degrades ETTR under frequent failures, falling to just 0.134 for DeepSeek‑MoE at MTBF=10M. \NAME consistently maintains high ETTR (0.951–0.964 at MTBF=10M) by effectively combining high frequency checkpointing with low runtime overhead. This approach allows \NAME to match MoC's fast recovery, CheckFreq and Gemini's low overhead without inheriting their respective penalties.

\begin{figure*}[t!]
    \centering
    \includegraphics[width=1\linewidth]{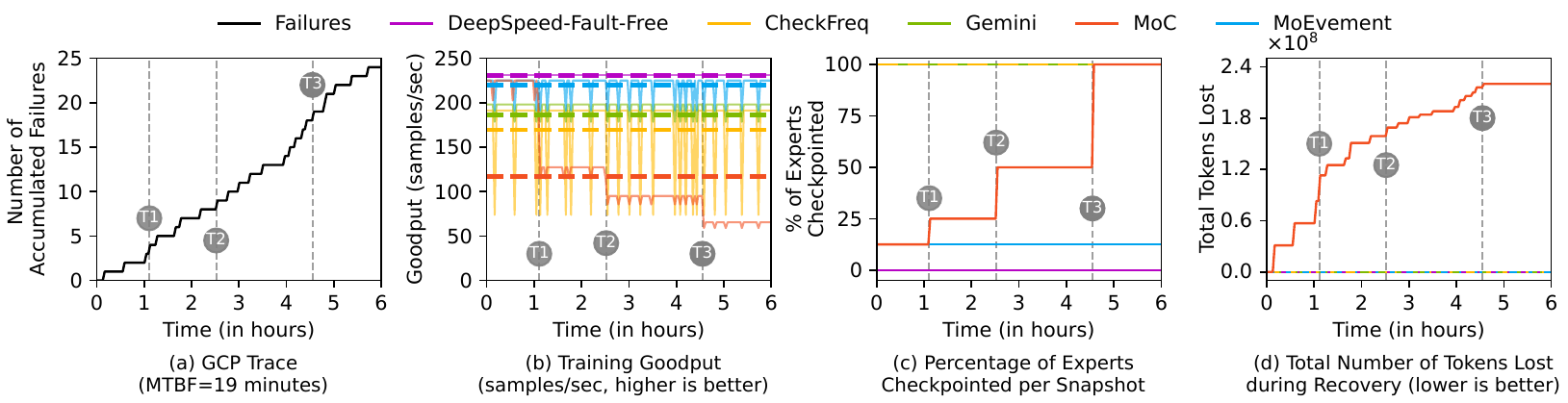}
    \vspace{-0.25in}
    \caption{Training DeepSeek‑MoE under a 6‑hour real-world failure trace. \NAME sustains the highest goodput throughout, while MoC's goodput declines as its lost‑token budget is exhausted, prompting more experts to be checkpointed per snapshot. CheckFreq and Gemini achieve stable but lower goodput due to long checkpoint intervals. In~\hyperref[fig:real-world-trace]{\ref*{fig:real-world-trace}b}, dashed lines show the average goodput over the 6‑hour trace.}
    \label{fig:real-world-trace}
    \vspace{-0.15in}
\end{figure*}

Ultimately, this superior training efficiency translates to tangible end-to-end performance improvements: at MTBF=10M, \NAME completes training up to $1.4\times$ faster than CheckFreq, $1.3\times$ faster than Gemini, and $7.1\times$ faster than MoC.

\subsection{Training Efficiency Under Dynamic Failures}\label{sec:real-world-trace}

To further evaluate \NAME under realistic conditions, we replay a 6-hour failure trace collected from Google Cloud Platform (GCP) instances, as used in recent works~\cite{bamboo,oobleck,recycle}. Figure~\hyperref[fig:real-world-trace]{\ref*{fig:real-world-trace}a} shows the 24 failure events in this period, corresponding to an average MTBF of $\approx19$ minutes.

Figure~\hyperref[fig:real-world-trace]{\ref*{fig:real-world-trace}b} compares goodput—useful throughput (samples/sec) excluding recomputed samples after failures—for each checkpointing method over the replayed trace. CheckFreq and Gemini exhibit consistently lower goodput due to frequent recomputation caused by their long checkpoint intervals (124 and 92 iterations, respectively). MoC initially achieves high goodput by checkpointing a smaller set of experts, as seen in Fig.~\hyperref[fig:real-world-trace]{\ref*{fig:real-world-trace}c}. However, as lost tokens from recovery with partial snapshots accumulate (Fig.~\hyperref[fig:real-world-trace]{\ref*{fig:real-world-trace}d}), MoC must progressively checkpoint a larger fraction of experts (from 12.5\% at \greycircledtikz{T1} to 100\% at \greycircledtikz{T3} in Fig.~\hyperref[fig:real-world-trace]{\ref*{fig:real-world-trace}c}) to preserve accuracy, which reduces its goodput by $71\%$ over time.

In contrast, \NAME maintains the highest goodput throughout the trace, delivering 1.25$\times$, 1.15$\times$, and 1.98$\times$ higher goodput than CheckFreq, Gemini, and MoC‑System, respectively. This demonstrates \NAME's effective balance of low checkpoint overhead, fast localized recovery, and no lost tokens under dynamic failure conditions.

\subsection{\NAME Scalability}\label{sec:scalability}

\begin{table}[!t]
  \centering
  \resizebox{0.85\linewidth}{!}{
  \begin{tabular}{c|c|ccc}
  \toprule
   \textbf{Model} & \textbf{System} & \textbf{1H} & \textbf{30M} & \textbf{10M} \\ \midrule
   \multirow{2}{*}{QWen-MoE}    &  Gemini   & $-0.14\%$    & $-1.47\%$    & $+1.02\%$ \\
                                &  \NAME    & $+0.86\%$    & $-0.73\%$    & $+0.30\%$ \\ \midrule
   \multirow{2}{*}{DeepSeek-MoE}&  Gemini   & $-0.87\%$    & $-0.56\%$    & $-0.19\%$ \\
                                &  \NAME    & $-0.59\%$    & $+1.22\%$    & $+0.25\%$ \\  
   \bottomrule
  \end{tabular}
  }
  \caption{Difference between simulated and measured ETTR.}
  \label{tbl:simulator-validation}
\end{table}

{\bf Simulator.} Due to the unavailability of a cluster with thousands of GPUs, we built a simulator to estimate Effective Training Time Ratio (ETTR) for arbitrary model and cluster configurations, given a specified MTBF and checkpointing technique (detailed in Appendix~\ref{appendix:simulator}). It leverages real-world profiled statistics for each pipeline operation, capturing both computation and communication costs. We validated accuracy against measurements on Azure for two MoE models under diverse failure scenarios. Table~\ref{tbl:simulator-validation} shows a maximum deviation of $1.47\%$, primarily due to minor runtime variations in NCCL collectives, which minimally affect \NAME's relative performance and scalability trends.

\begin{figure}[t!] 
    \centering
    \vspace{-0.1in}
    \includegraphics[width=1\linewidth]{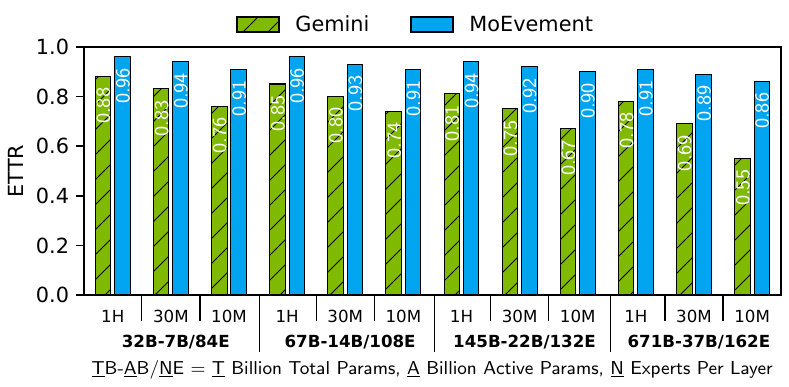}
    \vspace{-0.25in}
    \caption{Simulated ETTR of \NAME as model size increases.}
    \label{fig:scaling}
\end{figure}

\textit{\textbf{How effectively does \NAME scale to large models?}} Using the simulator, we evaluate DeepSeek-MoE models with 32B, 67B, 145B, and 671B total parameters across clusters configured as follows: (512 GPUs, 16 stages per pipeline, 4 pipelines), (1536 GPUs, 24 stages per pipeline, 8 pipelines), (4096 GPUs, 32 stages per pipeline, 16 pipelines), and (16384 GPUs, 64 stages per pipeline, 32 pipelines). All models employ 8-way expert parallelism, corresponding to the NVLink domain size. Figure~\ref{fig:scaling} presents the ETTR achieved by Gemini and \NAME across varying MTBF. 

\NAME consistently sustains high ETTR across increasingly large models and clusters, even as failure rates increase. At a 1-hour MTBF, \NAME achieves ETTRs above $0.91$ across all scales. As MTBF drops to 30 minutes and 10 minutes, ETTR declines gradually, but \NAME remains significantly more resilient than Gemini. For instance, on the 671B model, \NAME achieves ETTR of $0.86$ under 10-minute MTBF, compared to Gemini’s $0.55$; delivering $1.55\times$ faster training. This performance gap stems from fundamental differences in design. Gemini checkpoints less frequently and must roll back all machines—faulty and healthy—on failure. As pipeline depth and cluster size grow, the cost of global rollback becomes increasingly severe, compounding runtime disruption. In contrast, \NAME's localized recovery avoids full-cluster rollbacks and supports frequent checkpointing. This design allows \NAME to scale more gracefully under high failure rates, maintaining high training efficiency even at industrial scale.

\subsection{Impact of Failures on Model Accuracy}\label{sec:model-accuracy}

\begin{figure}[t!]
    \centering
    \includegraphics[width=\linewidth]{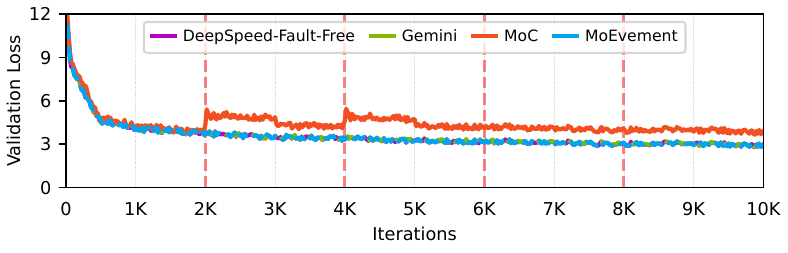}
    \vspace{-0.28in}
    \caption{Validation loss for DeepSeek-MoE during 10K iterations of training, with failures injected at 2K, 4K, 6K, 8K iterations.}
    \label{fig:model-accuracy}
\end{figure}

To evaluate how checkpointing and recovery affect model quality, we train DeepSeek‑MoE for 10,000 iterations, injecting faults at iterations 2K, 4K, 6K, and 8K (marked by red vertical lines in Fig.~\ref{fig:model-accuracy}). The validation loss trajectories for Gemini, MoC, \NAME, and a fault‑free DeepSpeed baseline reveal that MoC exhibits validation loss spikes after the first two failures due to token loss from partial recovery. After iteration 4K, MoC begins checkpointing all experts to prevent further degradation, avoiding new spikes but never regaining fault‑free performance. Gemini and \NAME closely track the fault-free DeepSpeed baseline, demonstrating that both effectively preserve model quality despite repeated failures.

\begin{table}[t!]
        \centering
        \resizebox{\linewidth}{!}{
        \begin{tabular}{@{}c|c|c|c|c|c@{}}
        \toprule
        \textbf{Task} &\textbf{\#Shot} & \textbf{\makecell{DeepSpeed\\Fault-Free}} & \textbf{Gemini} & \textbf{MoC} & \textbf{\NAME} \\ \midrule
        PIQA~\cite{piqa} & 0-shot & 72.4 & \Best{72.5} & \Worst{61.2} & 72.4 \\ \thinline
        HellaSwag~\cite{hellaswag} & 0-shot & \Best{68.9} & 68.8 & \Worst{53.1} & 68.8 \\ \thinline
        TriviaQA~\cite{triviaqa} & 5-shot & 54.8 & 54.6 & \Worst{37.5} & \Best{55.1} \\ \thinline
        NaturalQuestions~\cite{naturalquestions} & 5-shot & \Best{15.3} & 15.1 & \Worst{6.3} & \Best{15.3} \\ 
        \bottomrule
        \end{tabular}
        }
        \vspace{-0.075in} 
        \caption{Downstream evaluation on commonsense reasoning and knowledge-intensive tasks. We compare fault-free DeepSpeed (baseline) against Gemini, MoC, and \NAME for DeepSeek-MoE. \Best{Best} and \xspace\Worst{worst} results highlighted in green and red, respectively.}
        \label{tbl:downstream-eval}
\end{table}

We further assess downstream performance (on a 0–100 scale, higher is better) on PIQA~\cite{piqa} and HellaSwag~\cite{hellaswag} (commonsense reasoning) and TriviaQA~\cite{triviaqa} and NaturalQuestions~\cite{naturalquestions} (knowledge‑intensive QA) as shown in Table~\ref{tbl:downstream-eval}. \NAME achieves comparable accuracy to fault‑free DeepSpeed and Gemini baselines across all tasks, confirming no impact on downstream model quality. MoC, however, consistently under performs, highlighting the negative consequences of partial recovery and reinforcing the importance of preserving synchronous training semantics.

\subsection{\NAMESC Performance Breakdown}\label{sec:technique-breakdown}

\begin{figure}[t!] 
    \centering
    \vspace{-0.1in}
    \includegraphics[width=1\linewidth]{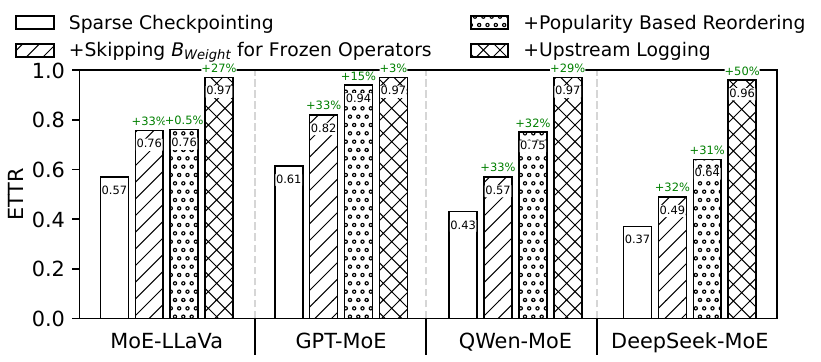}
    \caption{Incremental impact of \NAME's techniques across models. Each addition reduces recovery overhead, improving ETTR.}
    \label{fig:performance-breakdown}
\end{figure}

\textit{\textbf{What is each \NAME technique’s contribution to performance?}} We evaluate the incremental impact of \emph{sparse checkpointing}, \emph{skipping  $B_{\text{weight}}$ for frozen operators}, \emph{popularity based reordering}, and \emph{upstream logging} on training efficiency (ETTR) of \NAME. Figure~\ref{fig:performance-breakdown} reports ETTR for four models as each technique is added; green annotations show the relative improvement at each step.

\emph{Sparse checkpointing} establishes the baseline, yielding ETTRs of 0.57 (MoE-LLaVa), 0.43 (GPT-MoE), 0.37 (QWen-MoE), and 0.49 (DeepSeek-MoE). \emph{Skipping $B_{\text{weight}}$ and optimizer updates for frozen operators}—those lacking FP32 state—reduces recovery cost during sparse-to-dense conversion by $\approx33\%$, as their role is limited to forward and input-gradient propagation. \emph{Popularity-based reordering} defers the checkpointing of frequently activated experts, keeping them frozen longer and extending the compute savings. Its impact grows with expert count: models like QWen-MoE and DeepSeek-MoE (64 experts) see ETTR improve by $32\%$, GPT-MoE (32 experts) by $15\%$, while MoE-LLaVa (4 experts) sees no change. With more experts, the checkpointing order has greater influence over how long operators remain frozen, making reordering more effective. Finally, \emph{upstream logging} confines recovery to the failed stage’s data-parallel group by logging activations and gradients in host memory. This reduces rollback scope and eliminates pipeline bubbles, raising ETTR to $\approx$0.97 across all models. The largest gain ($+50\%$) is seen in DeepSeek-MoE due to its 12-stage pipeline, while the smallest improvement is in GPT-MoE, whose shallow pipeline limits the benefit of localized recovery.

\begin{table}[!t]
  \vspace{-0.125in} 
  \centering
  \resizebox{\linewidth}{!}{
  \begin{tabular}{c|cc|cc|c}
  \toprule
    \multirow{2}{*}{\textbf{Model}} & \multicolumn{2}{c|}{\textbf{Gemini}} & \multicolumn{2}{c|}{\textbf{\NAME}} & \multirow{2}{*}{\textbf{\makecell{Increase over \\Gemini (\%)}}} \\ \cmidrule(lr){2-3} \cmidrule(lr){4-5}
  & \textbf{GPU} & \textbf{CPU} & \textbf{GPU} & \textbf{CPU (X + Y)} &  \\ \midrule 
  MoE-LLaVa     & $0$     & $75.4$      & $0$     & $83.1~(81.2 + 1.9)$    & $+10.1\%$  \\ \thinline
  GPT-MoE       & $0$     & $189.8$     & $0$     & $209.4~(205.2 + 4.2)$   & $+10.3\%$  \\ \thinline
  QWen-MoE      & $0$     & $371.6$     & $0$     & $433.2~(420.8 + 12.4)$   & $+16.5\%$  \\ \thinline
  DeepSeek-MoE  & $0$     & $426.4$     & $0$     & $499.8~(478.7 + 21.1)$  & $+17.2\%$  \\ 
  \bottomrule
  \end{tabular}
  }
  \vspace{-0.075in} 
  \caption{Memory footprint (in GB) of Gemini and \NAME on GPU and CPU. For \NAME, CPU memory footprint is decomposed as $X + Y$, with $X$ representing the sparse checkpoint size (in GB) and $Y$ representing the activation and gradient log size (in GB).}
  \label{tbl:memory-overhead}
\end{table}

\begin{table*}[t!]
  \centering
  \resizebox{\textwidth}{!}{%
  \begin{tabular}{ccc|cccc|cccc|c|cccc|cccc}
  \toprule
  \multicolumn{3}{c|}{\textbf{Training Configuration}} 
  & \multicolumn{4}{c|}{\textbf{Checkpointing Interval}} 
  & \multicolumn{4}{c|}{\textbf{Avg. Per-Iteration Checkpointing Overhead}} 
  & \multirow{3}{*}{\textbf{MTBF}} 
  & \multicolumn{4}{c|}{\textbf{Total Recovery Time}} 
  & \multicolumn{4}{c}{\textbf{Effective Training Time Ratio (ETTR)}} \\ 
  
  & & & \multicolumn{4}{c|}{\textbf{(iterations)}} 
  & \multicolumn{4}{c|}{\textbf{(seconds, overhead \% in parentheses)}} 
  &  
  & \multicolumn{4}{c|}{\textbf{(seconds)}} 
  & \multicolumn{4}{c}{
    \raisebox{-0.65\height}{
    \begin{tikzpicture}
    \begin{axis}[
        hide axis,
        scale only axis,
        height=0.2cm,
        width=3.5cm,
        colorbar horizontal,
        point meta min=0.0,
        point meta max=1.0,
        colorbar style={
            width=3.5cm,
            xtick={0.0, 0.25, 0.5, 0.75, 1.0},
            xticklabel style={font=\small},
            height=0.2cm,
        },
        colormap={redyellowgreen}{
            rgb(0pt)=(1,0,0)           
            rgb(83.33pt)=(1,1,0)       
            rgb(100pt)=(0,1,0)         
        },
        colorbar sampled,
    ]
    \addplot [draw=none] coordinates {(0,0)};
    \end{axis}
    \end{tikzpicture}
    }
  } \\ \cline{1-3} \cline{4-11} \cline{13-20}
   \textbf{\makecell{Compute \\ Weight}} & \textbf{\makecell{Master \\ Weight}} & \textbf{\makecell{Optimizer \\ State}}
   & \textbf{CheckFreq} & \textbf{Gemini} & \textbf{MoC} & \textbf{\NAME} 
   & \textbf{CheckFreq} & \textbf{Gemini} & \textbf{MoC} & \textbf{\NAME} & 
   & \textbf{CheckFreq} & \textbf{Gemini} & \textbf{MoC} & \textbf{\NAME} 
   & \textbf{CheckFreq} & \textbf{Gemini} & \textbf{MoC} & \textbf{\NAME} \\ \hline
   
   \multirow{3}{*}{FP16} & \multirow{3}{*}{FP16} & \multirow{3}{*}{FP16 + FP16~\cite{collage}}
    & \multirow{3}{*}{77} & 71 & \multirow{3}{*}{1} & \multirow{3}{*}{\makecell{1\\ $(W_\text{sparse}=3)$}} 
    & \multirow{3}{*}{0.10 (3\%)} & 0.11 (3\%) & 1.16 (39\%) & \multirow{3}{*}{0.06 (2\%)} 
    & 1H 
        & 2036 & 1854 & 79 & 86 
        & \ApplyGradient{0.918} & \ApplyGradient{0.925} 
        & \ApplyGradient{0.613} & \ApplyGradient{0.984} \\
    & & & & 48 & & & & 0.16 (6\%) & 2.91 (97\%) & & 30M 
        & 3549 & 2471 & 102 & 113 
        & \ApplyGradient{0.876} & \ApplyGradient{0.886} 
        & \ApplyGradient{0.459} & \ApplyGradient{0.979} \\
    & & & & 27 & & & & 0.29 (10\%) & 3.51 (117\%) & & 10M 
        & 10648 & 4324 & 246 & 268
        & \ApplyGradient{0.706} & \ApplyGradient{0.795} 
        & \ApplyGradient{0.425} & \ApplyGradient{0.973} \\ \hline

   \multirow{3}{*}{FP8} & \multirow{3}{*}{FP32} & \multirow{3}{*}{FP32 + FP32~\cite{micikevicius2022}}
    & \multirow{3}{*}{227} & 182 & \multirow{3}{*}{1} & \multirow{3}{*}{\makecell{1\\ $(W_\text{sparse}=6)$}} 
    & \multirow{3}{*}{0.06 (3\%)} & 0.09 (4\%) & 3.53 (154\%) & \multirow{3}{*}{0.05 (2\%)} 
    & 1H 
        & 5162 & 3591 & 48 & 59 
        & \ApplyGradient{0.853} & \ApplyGradient{0.878} 
        & \ApplyGradient{0.357} & \ApplyGradient{0.952} \\
    & & & & 94 & & & & 0.17 (8\%) & 8.89 (386\%) & & 30M 
        & 9524 & 3709 & 75 & 96 
        & \ApplyGradient{0.748} & \ApplyGradient{0.839} 
        & \ApplyGradient{0.196} & \ApplyGradient{0.946} \\
    & & & & 46 & & & & 0.34 (15\%) & 10.70 (465\%) & & 10M 
        & 28574 & 5446 & 185 & 232 
        & \ApplyGradient{0.302} & \ApplyGradient{0.718} 
        & \ApplyGradient{0.115} & \ApplyGradient{0.941} \\ \hline

   \multirow{3}{*}{FP8} & \multirow{3}{*}{FP16} & \multirow{3}{*}{FP32 + FP32~\cite{mellempudi2019mixed}}
    & \multirow{3}{*}{205} & 157 & \multirow{3}{*}{1} & \multirow{3}{*}{\makecell{1\\ $(W_\text{sparse}=4)$}} 
    & \multirow{3}{*}{0.06 (3\%)} & 0.08 (4\%) & 2.88 (137\%) & \multirow{3}{*}{0.04 (2\%)} 
    & 1H 
        & 3763 & 2846 & 45 & 53 
        & \ApplyGradient{0.881} & \ApplyGradient{0.893} 
        & \ApplyGradient{0.368} & \ApplyGradient{0.966} \\
    & & & & 79 & & & & 0.16 (8\%) & 7.25 (345\%) & & 30M 
        & 7527 & 2882 & 71 & 83 
        & \ApplyGradient{0.793} & \ApplyGradient{0.852} 
        & \ApplyGradient{0.226} & \ApplyGradient{0.958} \\
    & & & & 43 & & & & 0.30 (14\%) & 8.73 (416\%) & & 10M 
        & 22581 & 4756 & 151 & 172 
        & \ApplyGradient{0.439} & \ApplyGradient{0.738} 
        & \ApplyGradient{0.163} & \ApplyGradient{0.953} \\ \hline

   \multirow{3}{*}{FP8} & \multirow{3}{*}{FP16} & \multirow{3}{*}{FP8 + FP16~\cite{peng2023}}
    & \multirow{3}{*}{94} & 68 & \multirow{3}{*}{1} & \multirow{3}{*}{\makecell{1\\ $(W_\text{sparse}=3)$}}
    & \multirow{3}{*}{0.07 (3\%)} & 0.10 (5\%) & 1.16 (61\%) & \multirow{3}{*}{0.04 (2\%)} 
    & 1H 
        & 1554 & 1108 & 52 & 57 
        & \ApplyGradient{0.916} & \ApplyGradient{0.923} 
        & \ApplyGradient{0.557} & \ApplyGradient{0.976} \\
    & & & & 61 & & & & 0.11 (6\%) & 2.92 (154\%) & & 30M 
        & 3107 & 2022 & 85 & 94 
        & \ApplyGradient{0.889} & \ApplyGradient{0.894} 
        & \ApplyGradient{0.349} & \ApplyGradient{0.968} \\
    & & & & 29 & & & & 0.22 (12\%) & 3.52 (185\%) & & 10M 
        & 9322 & 2934 & 196 & 213 
        & \ApplyGradient{0.739} & \ApplyGradient{0.806} 
        & \ApplyGradient{0.321} & \ApplyGradient{0.965} \\ \hline

   \multirow{3}{*}{FP8} & \multirow{3}{*}{FP8} & \multirow{3}{*}{FP8 + FP16~\cite{peng2023}}
    & \multirow{3}{*}{78} & 59 & \multirow{3}{*}{1} & \multirow{3}{*}{\makecell{1\\ $(W_\text{sparse}=3)$}} 
    & \multirow{3}{*}{0.07 (3\%)} & 0.09 (5\%) & 0.87 (51\%) & \multirow{3}{*}{0.03 (1\%)} 
    & 1H 
        & 1153 & 876 & 40 & 44 
        & \ApplyGradient{0.937} & \ApplyGradient{0.943} 
        & \ApplyGradient{0.584} & \ApplyGradient{0.979} \\
    & & & & 52 & & & & 0.18 (10\%) & 2.19 (129\%) & & 30M 
        & 2307 & 1516 & 79 & 85 
        & \ApplyGradient{0.905} & \ApplyGradient{0.907} 
        & \ApplyGradient{0.381} & \ApplyGradient{0.976} \\
    & & & & 24 & & & & 0.31 (17\%) & 2.63 (155\%) & & 10M 
        & 6921 & 2100 & 162 & 178 
        & \ApplyGradient{0.752} & \ApplyGradient{0.816} 
        & \ApplyGradient{0.326} & \ApplyGradient{0.971} \\
   \bottomrule
  \end{tabular}%

  }
  \vspace{-0.075in} 
  \caption{Comparison of checkpointing techniques across low precision training configurations for DeepSeek-MoE over a 12-hour run.}
  \label{tbl:low-precision-training}
  \vspace{-0.125in} 
\end{table*}

\textit{\textbf{How much extra memory does \NAME use?}} \NAME adds \emph{no} GPU memory overhead; all additional state lives in CPU memory, consistent with Gemini. The extra CPU memory usage arises from two sources: (\emph{i}) FP16 compute weights in sparse checkpoint (\(X\)) for \emph{frozen} operators awaiting their full FP32 training state, and (\emph{ii}) activation and gradient logs (\(Y\)) for localized recovery. Across the evaluated models in Table~\ref{tbl:memory-overhead}, the increase is at most \(17.2\%\) relative to Gemini's dense checkpoint (and similarly for CheckFreq, omitted for brevity). This modest increase, \(\leq 2\%\) of the available 10\,TB CPU memory, enables \NAME to sustain high ETTR without compromising synchronous training semantics.

\subsection{Low-Precision Training Beyond FP16 \& FP32}\label{sec:low-precision-exp}

Low-precision training has emerged as a promising approach for scaling next-generation models by accelerating iterations, reducing memory footprint, and lowering communication overheads~\cite{lee2025fp8againquantifyingreduced}. Recent GPUs such as Nvidia H100 introduce native support for FP8 datatypes~\cite{h100-supports-fp8}, making low-precision training increasingly practical at scale. To evaluate checkpointing performance in this regime, we use a private cluster with 16 nodes, each equipped with a 104-core Intel Xeon processor, 2.1 TB of RAM, and eight Nvidia H100 80 GB GPUs. Within a node, GPUs are connected by 900 GB/s NVLink, and across nodes by 200 Gbps InfiniBand. We evaluate all four systems across five low-precision configurations proposed in prior work~\cite{collage, micikevicius2022, mellempudi2019mixed, peng2023}, training DeepSeek-MoE (Table~\ref{tbl:model-spec}) with 8-way pipeline parallelism, 2-way data parallelism, and 8-way expert parallelism, using a batch size of 512 (micro-batch size 32) and a sequence length of 2048 tokens.

The five configurations vary precision used for computations, master weights, and optimizer state. The choice of precision impacts two key variables: iteration time and snapshot size. Switching compute from FP16 to FP8 shortens iterations, shrinking the window to overlap snapshot I/O. Lowering the precision of training state (e.g., FP8 master weights and FP8+FP16 optimizer state) reduces the snapshot size by as much as 66\%. Dense baselines are limited by the tighter of these two constraints: shorter iterations force longer intervals, while smaller snapshots from lower-precision states enable shorter intervals. For CheckFreq, we configure its policy module to choose intervals that maintain runtime overhead below 3\%. For Gemini, we apply an oracle policy: for each MTBF, we sweep intervals offline and select the one maximizing ETTR; this hindsight-informed choice requires knowledge unavailable at runtime, thus upper-bounding Gemini’s achievable performance. MoC and \NAME both checkpoint every iteration, albeit selectively.

As shown in Table~\ref{tbl:low-precision-training}, CheckFreq's shortest feasible interval (78 iterations) occurs when training state uses FP8 master weights and FP8+FP16 optimizer states. When using FP32 optimizer state with FP16 or FP32 model weights, CheckFreq is forced to lengthen intervals significantly, to 205 and 227 iterations respectively. Gemini follows a similar trend at MTBF=1H, increasing its interval from 59 to 182 as state precision rises; under frequent failures its oracle policy trades higher runtime overhead for shorter rollbacks, checkpointing every 24-94 iterations at 6-13\% per-iteration overhead. MoC's overhead grows with failure frequency because it exhausts its lost‑token budget sooner and is forced to increase experts checkpointed per iteration—eventually checkpointing all experts—and incurring 39--465\% per-iteration overhead. In contrast, \NAME\ maintains 1–2\% overhead across all configurations with small sparse windows.

These interval choices directly shape recovery costs and ETTR. At higher precision (FP32), dense baselines suffer from long intervals and high recovery costs. At MTBF=1H, CheckFreq's recovery time ranges from 1153 to 5162 secs, and Gemini’s from 876 to 3591 secs. At MTBF=10M, these costs increase to 6921–28574 secs (CheckFreq) and 2100–5446 secs (Gemini), even when Gemini selects intervals using an oracle. As precision decreases (FP8 master weights with FP8+FP16 optimizer state), dense baselines improve: Gemini’s ETTR at MTBF=10M rises from 0.718 to 0.816. However, CheckFreq and Gemini still sacrifice 10–25\% ETTR under frequent failures. MoC consistently under performs across all precisions, with ETTR dropping from 0.36–0.61 at MTBF=1H to 0.12–0.43 at MTBF=10M. In contrast, \NAME maintains low and stable overhead (1–2\%), consistently achieving ETTR of 0.94–0.98 across all precisions and MTBFs. Consequently, \NAME provides up to an 8$\times$ improvement in end-to-end training time.
\section{Conclusion}\label{sec:conclusion}

We presented \NAME, a distributed, in-memory checkpointing system designed for efficient and reliable MoE model training. \NAME introduces three key techniques—\emph{sparse checkpointing}, \emph{sparse-to-dense checkpoint conversion}, and \emph{lightweight upstream activation and gradient logging}—which together break the runtime–recovery tradeoff, mitigate the correctness–efficiency tension, and eliminate excessive global recomputation in prior approaches. Our evaluation shows that \NAME sustains ETTR $\ge 0.94$ even at MTBFs as low as 10 minutes, achieving up to an $8\times$ end-to-end training speedup without compromising model accuracy.

\section*{Acknowledgements}
We are grateful to the anonymous reviewers and to our shepherd, Dan Li, whose comments have greatly helped improve this paper. This research was partly supported by the Stanford Platform Lab and its affiliates, and by ACE, one of the seven centers in JUMP 2.0, a Semiconductor Research Corporation (SRC) program sponsored by DARPA. 
\vspace{0.1in}



\bibliographystyle{plain}
\bibliography{references}

\clearpage      
\nobalance      
\appendix

\section{Recovery from Concurrent Failures}\label{appendix:concurrent-failures}

Concurrent failures during distributed training can occur in two scenarios: \emph{multiple simultaneous failures}, where two or more workers fail concurrently, and \emph{cascading failures}, where additional failures occur during ongoing recovery operations.

\begin{figure}
    \centering
    \includegraphics[width=\linewidth]{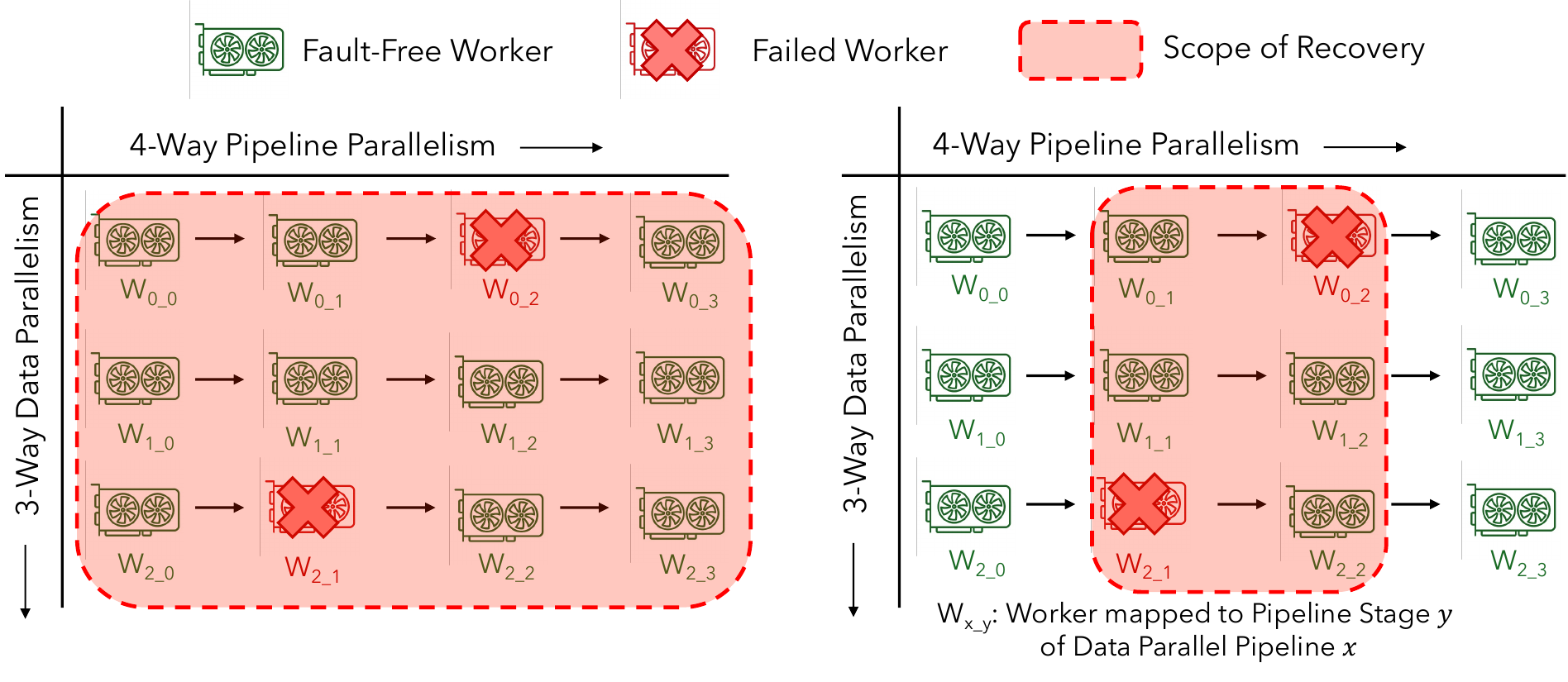}
    \vspace{-0.25in}
    \caption{Scope of recovery from multiple simultaneous failures \emph{without} (left) and \emph{with} (right) localized recovery.} 
    \label{fig:recovery-from-multiple-failures}
\end{figure}

\noindent \textbf{Multiple Simultaneous Failures.} Upon detecting a failure, \NAME pauses all workers and aborts the current iteration. Failed workers are promptly replaced with healthy spares. Only the data-parallel (DP) groups containing the failed workers roll back to their most recent sparse checkpoints, typically captured within the last $W_{\text{sparse}}$ iterations. Unaffected workers remain paused to maintain global consistency. If the failed workers form a contiguous pipeline segment  (Fig.~\hyperref[fig:recovery-from-multiple-failures]{\ref*{fig:recovery-from-multiple-failures}~(right)}), \NAME initiates a \emph{joint localized recovery}. During this joint recovery, boundary stages adjacent to the affected segment supply logged activations and gradients, enabling the impacted DP groups to collaboratively replay computations and convert their sparse checkpoints into a dense checkpoint. Conversely, if failures involve nonadjacent workers, each affected DP group independently and concurrently executes its own localized recovery. In both cases, overall recovery time is determined by the slowest individual or joint recovery, after which training resumes across all workers.

\noindent \textbf{Cascading Failures.} \NAME dynamically adapts the recovery scope in response to cascading failures. If a subsequent failure occurs in a worker that is adjacent to, or already part of, an ongoing recovery, \NAME expands the recovery scope to include this newly failed worker, forming an enlarged contiguous segment, and restarts \emph{joint localized recovery} for the affected DP groups. During recovery, each DP group rolls back to the most recent persisted sparse checkpoint common to that region, and replays computations by utilizing activation and gradient logs from healthy boundary neighbors. If the new failure is disjoint from ongoing recoveries, existing recoveries proceed without interruption, and the new failure triggers a separate, independent recovery.

\section{Alternative Operator Ordering Schemes}\label{appendix:ordering-scheme}

While \NAME defaults to ordering experts by ascending popularity (\ie activation count) within each sparse checkpoint window, the \texttt{OrderOperators()} interface supports alternative ordering schemes to accommodate different model and workload characteristics:

\noindent \textbf{Soft-Count Popularity.} Instead of using binary (hard) activation counts, popularity can be estimated more smoothly by aggregating gating probabilities. Let $\mathcal{P}_j^l\left(x_i\right)$ represent the gating probability assigned to expert $j$ in layer $l$ for token $x_i$, where $\mathcal{D}$ is the training dataset. Then, the soft-count popularity is defined as:
\[
\mathcal{A}_j^l = \sum_{x_i \in \mathcal{D}} \mathcal{P}_j^l\left(x_i\right)
\] 

\noindent \textbf{Time-Decayed Popularity.} To adaptively track popularity over changing activation patterns during training, we use an exponential moving average over recent mini-batches. Let $\mathcal{B}_t$ be the set of tokens in mini-batch $t$, and let $\mathbf{1}\left[\gamma(e_j^l, x_i) = 1\right]$ indicate whether expert $j$ in layer $l$ is activated for token $x_i$, where $\gamma(e_j^l, x_i)$ is a binary indicator function, and \(\alpha\) controls the rate of decay. The time-decayed popularity is:
\[
\mathcal{A}_j^l(t) = \alpha \times \mathcal{A}_j^l(t-1) + (1-\alpha) \times \sum_{x_i \in \mathcal{B}_t} \mathbf{1}\left[\gamma(e_j^l, x_i) = 1\right]
\]  

\noindent \textbf{Capacity-Aware Ordering.} For heterogeneous experts~\cite{hmoe, tcmoe} (\eg with different capacity factors), ordering can be weighted by utilization relative to capacity, prioritizing those with lower relative utilization. If each expert $j$ in layer $l$ has a capacity factor $\mathcal{C}^l_j$ (maximum tokens it can process per batch), we define a capacity-normalized popularity score:
\[
\hat{\mathcal{A}}^{l}_{j} = \frac{\mathcal{A}_j^l}{\mathcal{C}^l_j}
\] 

\section{Simulator for Estimating ETTR}\label{appendix:simulator}

We implement a simulator that estimates iteration time and Effective Training Time Ratio (ETTR) given a model architecture, cluster configuration, parallelization plan, and Mean Time Between Failures (MTBF). We validate its accuracy against measurements collected from Azure using two MoE models under diverse failure scenarios (Table~\ref{tbl:simulator-validation}). Validation results show a maximum deviation of 1.47\%, primarily due to runtime variability in NCCL operations. These small variations have negligible impact on observed relative performance trends, demonstrating the simulator’s effectiveness in evaluating distributed training performance without requiring direct access to large-scale GPU clusters. We describe its implementation below:

\noindent\textbf{Inputs.} Our simulator requires four primary inputs: (1) a training job specification, including the model architecture, \texttt{global\_batch\_size}, optimizer configuration, and hyper-parameters such as \texttt{micro\_batch\_size}; (2) a cluster configuration specifying GPU type and count, GPU allocation per node, inter-node bandwidth, topology, and hardware heterogeneity; and (3) a parallelization plan describing pipeline stage partitioning, tensor- and data-parallel degrees, and MoE expert placement or sharding, where applicable; and (4) a Mean Time Between Failures (MTBF) to characterize failure rates and corresponding recovery overhead.

\noindent\textbf{Profiler.} We derive computational and communication costs from empirical profiling of real training runs. The profiler captures compute and memory requirements by executing the training job on a single GPU node for each GPU node type available in the resource pool. To minimize profiling overhead, repeated layers are represented by a single instance (\eg one transformer layer for large language models). We employ PyTorch hooks in DeepSpeed to measure the forward pass, backward pass, and parameter update times for varying micro batch sizes and tensor parallel degrees, using CUDA events to ensure accurate GPU timing measurements. Additionally, the profiler records parameters count, output activation sizes, and intermediate memory consumption per layer using PyTorch's CUDA memory allocator. Profiling overhead is minimal, typically completing in a few minutes.

For cluster-level profiling, the simulator gathers network bandwidth data between pairs of machine types. Since network bandwidth varies with message size, we measure bandwidth using PyTorch collectives with the NCCL backend across multiple message sizes, fitting a polynomial function to derive bandwidth coefficients for each node-type pair. NCCL collective operations use an affine model:
$$
T_{\text{NCCL}}(m,p) = \alpha(p) + \beta(p) \times m
$$

where $m$ is message size, and $p$ is group size. This captures NCCL-specific algorithmic behavior and network characteristics.

\noindent\textbf{Estimating Iteration-time.} Each iteration runs forward and backward passes over the global batch and then applies an optimizer update. We partition the global batch $B$ into $M$ micro batches and execute them with pipeline parallelism using an interleaved 1F1B schedule that proceeds through warm-up, steady state, and cool-down phases. Let $S$ be the number of pipeline stages, and let $t_s$ denote the measured per-micro-batch time for stage $s$ (forward, backward, and local communication). For a single data-parallel pipeline, the time spent on the forward and backward passes is:
$$
T_{\text{pipeline}} = (M + S - 1) \times \max_{s \in [1,S]} \left(t_s\right)
$$

To estimate iteration time, we combine pipeline execution time with global synchronization overhead from all-reduce and optimizer update time:
$$
T_{\text{iter}} = \max_{i \in [1,PP]} \left({T^{i}_\text{pipeline}} \right) + T_{\text{sync}} + T_{\text{update}}
$$

where $max$ is taken over $PP$ data-parallel pipelines to account for potential stragglers due to network variability. We estimate $T_{\text{sync}}$ from empirically measured intra- and inter-node bandwidths, and we incorporate observed overlap between computation and communication rather than assuming full serialization. $T_{\text{update}}$ is empirically profiled and reflects the time needed to compute and apply parameter updates.

\noindent\textbf{Estimating ETTR.} ETTR quantifies training efficiency under failures by combining checkpointing overhead with recovery overhead. Modeling failures as a Poisson process with rate $\frac{1}{\text{MTBF}}$, ETTR is estimated as:
$$
\text{ETTR} = \frac{1}{1 + \frac{T_{\text{ckpt}}}{T_{\text{iter}} \times \text{Ckpt}_{\text{interval}}}} \times \frac{1}{1 + \frac{\mathbb{E}[R]}{\text{MTBF}}},
$$

Here, $\frac{T_{\text{ckpt}}}{T_{\text{iter}} \times \text{Ckpt}_{\text{interval}}}$ represents the runtime overhead incurred by checkpointing every $\text{Ckpt}_{\text{interval}}$ iterations, and $\mathbb{E}[R]\approx\frac{1}{2}\times\text{Ckpt}_{\text{interval}}\times T_{\text{iter}}$ is the expected recovery time per failure under a uniform failure distribution within a checkpoint interval. This formulation explicitly captures the trade-off involved in checkpoint interval selection: longer intervals lower runtime overhead but increase potential recomputation after failures, whereas shorter intervals reduce recovery overhead but incur higher runtime checkpointing costs.

\section{Effect of Varying Expert Popularity}\label{appendix:popularity-effect}

Mixture-of-Experts (MoE) models frequently exhibit skewed activation patterns, with some experts consistently receiving more tokens despite the use of auxiliary load-balancing loss, and this skew often varies over time and across layers~\cite{flexmoe, fastermoe, fastmoe, smartmoe}. To analyze how varying degrees of skewness in expert popularity affect \NAME's performance, we quantify skewness using the normalized Herfindahl–Hirschman Index (HHI)~\cite{herfindahl}, defined as follows:
\[
\text{HHI} = \sum_{i=1}^{E} p_i^2    
\quad \text{and} \quad 
\text{S} = \frac{\text{HHI}-\frac{1}{E}}{1-\frac{1}{E}}
\]

\begin{figure*}[t!]
    \centering
    \begin{minipage}[]{0.48\textwidth}
        \centering
        \includegraphics[width=1\linewidth]{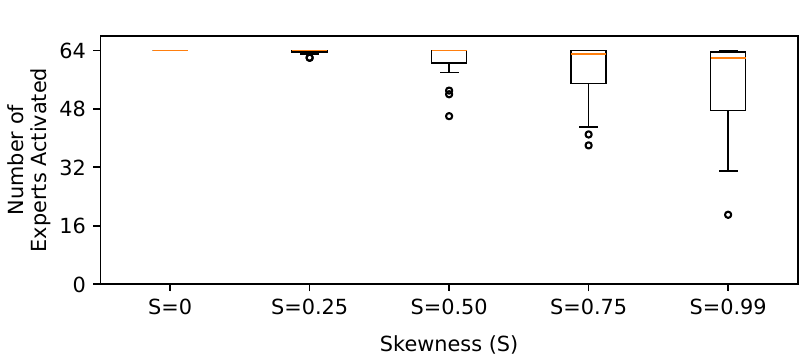}
        \vspace{-0.225in}
        \caption{Box plot showing the number of experts activated (assigned at least one token) per iteration across varying expert popularity skewness. Despite skewness concentrating tokens among fewer experts, most experts remain active. Boxes extend from the first quartile (Q1) to the third quartile (Q3), with a line at the median.}
        \label{fig:expert-box}
    \end{minipage}
    \hfill
    \begin{minipage}[]{0.48\textwidth}
        \centering
        \includegraphics[width=1\linewidth]{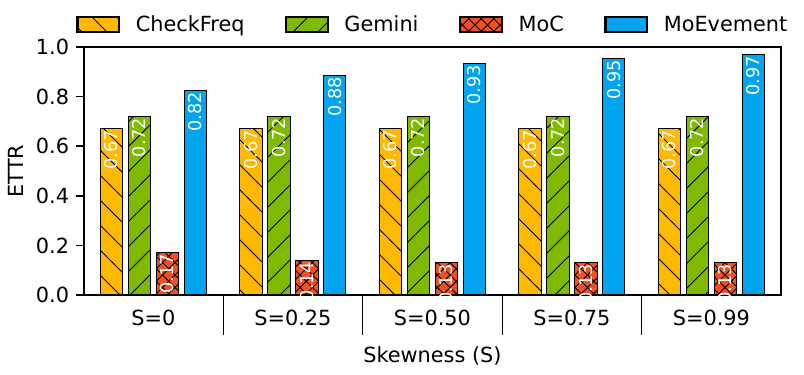}
        \vspace{-0.25in}
        \caption{Impact of Expert Popularity Skewness on ETTR across CheckFreq, Gemini, MoC, and \NAME using DeepSeek-MoE with varying levels of expert popularity skewness. Higher skewness enhances \NAME's advantage, as its popularity-based expert reordering reduces recovery overhead.}
        \label{fig:expert-diversity}
    \end{minipage}
    \vspace{-0.15in}
\end{figure*}

where $p = (p_1, \dots, p_E)$ represents the fraction of tokens assigned to each expert, with $\sum p_i = 1$, and $E$ is the number of experts ($E \ge 2$). The skewness metric $\text{S}$ ranges from $0$ for perfectly uniform popularity (each expert receives an equal share of tokens) to $1$ for maximum skewness (one expert receives all tokens). We systematically vary intermediate levels of skewness by sampling expert token distributions $p$ from a symmetric Dirichlet distribution~\cite{johnson1972continuous} parameterized by $\alpha$. Large values of $\alpha$ yield nearly uniform distributions, while small values yield highly skewed distributions. The expected HHI ($\mathbb{E}[\text{HHI}]$) and skewness ($\mathbb{E}[\text{S}]$) can be computed as:
\[
\mathbb{E}[\text{HHI}] = \frac{\alpha+1}{\alpha \times E + 1}
\quad \text{and} \quad 
\mathbb{E}[\text{S}] = \frac{\mathbb{E}[\text{HHI}]-\frac{1}{E}}{1-\frac{1}{E}}.
\] 

We report the Effective Training Time Ratio (ETTR) for CheckFreq, Gemini, MoC, and \NAME using DeepSeek-MoE with an MTBF of 10 minutes across varying skewness levels over a 12-hour long training run. All experiments use 96 A100 GPUs on Azure, employing standard FP16-FP32 mixed-precision~\cite{megatron-sc, gopher} training with the AdamW~\cite{adam} optimizer (additional details about the test cluster can be found in~\cref{sec:experimental-setup}). The target skewness levels $S\in\{0.25, 0.50, 0.75, 0.99\}$ correspond to $\alpha\approx\{0.0469, 0.0156, 0.0052, 0.000158\}$.

Figure~\ref{fig:expert-diversity} shows that $S=0$ (uniform popularity) represents the worst-case scenario for \NAME since uniform distribution of tokens across experts provides no clear popular subset of experts to defer, thereby limiting the benefits of popularity-based expert reordering. As $S$ increases to 0.25, 0.50, 0.75, and 0.99, \NAME's advantage over prior methods widens: higher skewness lets \NAME defer the most popular experts longer within each sparse window, which reduces recomputation during sparse‑to‑dense conversion and lowers recovery overhead. Importantly, despite increased skewness concentrating tokens among fewer experts, Fig.~\ref{fig:expert-box} demonstrates that majority of experts remain active, \ie they are assigned at least one token per iteration. This necessitates checkpointing all experts to avoid token loss upon failure, highlighting a crucial design consideration explicitly addressed by \NAME. Specifically, \NAME leverages popularity-based reordering to defer the most frequently activated experts, yet crucially ensures every expert, regardless of popularity, is checkpointed within each sparse window. This design preserves training consistency, completely eliminating any risk of token loss and thus maintaining equivalence to fault-free training semantics.

\NAME's design principle of zero token loss contrasts sharply with MoC, which inherently risks token loss due to its partial-checkpointing strategy. MoC checkpoints only a subset of experts per iteration in a round-robin fashion, making its ETTR sensitive to skewed expert popularity distributions. With increased skewness, a small subset of popular experts processes the majority of tokens. If a failure occurs after a popular expert has not been checkpointed for roughly $\frac{E}{K}$ iterations (where $E$ is the total number of experts, and $K$ is the number checkpointed per iteration), the resulting burst of token loss can rapidly exhaust MoC's token-loss budget. To avoid exceeding this budget, MoC is forced to checkpoint more experts per iteration, directly increasing per-iteration runtime overhead due to checkpoint-induced stalls and thereby lowering its ETTR. Empirically, as skewness \(S\) increases from 0 (uniform popularity) to 0.99 (highly skewed popularity), MoC's ETTR drops from 0.17 to 0.13 (Fig.~\ref{fig:expert-diversity}). 

CheckFreq and Gemini avoid popularity-based differentiation by checkpointing all experts every \(Ckpt_{\text{interval}}\) iterations. Consequently, their performance remains insensitive to variations in expert popularity skewness, resulting in stable but consistently lower ETTR (0.67 for CheckFreq and 0.72 for Gemini) compared to \NAME. Overall, \NAME outperforms prior approaches across all levels of skewness, achieving up to a $7.5\times$ reduction in end-to-end training time.

\section{Generalizing \NAME to Dense Models}\label{appendix:generalizing-to-dense}

Sparse checkpointing in \NAME leverages architectural properties unique to MoE models—each expert can be treated as an independent checkpointable unit, and the distribution of tokens across experts is both dynamic and inherently skewed (Fig.~\ref{fig:expert-popularity}). Generalizing sparse checkpointing techniques to dense transformer models represents a compelling direction for future research. Dense transformer architectures typically consist of monolithic feed-forward layers, which cannot be readily partitioned at a finer granularity akin to MoE experts. Nevertheless, each transformer layer itself can be considered an independently checkpointable unit. Sparse checkpointing could therefore be adapted to dense models by incrementally checkpointing subsets of consecutive layers across multiple iterations. Given the inherent directional flow of activations (forward) and gradients (backward), selecting subsets of layers starting from the output (back layers) and progressing toward the input (front layers) can strategically reduce recomputation costs during recovery.

Moreover, the localized recovery mechanism in \NAME, facilitated by upstream logging (\cref{sec:upstream-logging}), is directly applicable to dense models. Localized recovery can confine recomputation exclusively to affected data-parallel groups, thereby eliminating pipeline bubbles that typically occur during warm-up and cool-down phases of recovery. This advantage may become increasingly pronounced in training configurations with deep pipelines, which are common in large-scale transformer-based models. Quantifying and optimizing these benefits within dense model architectures represents an exciting and promising area for future work.

\end{document}